\documentclass[sigconf]{acmart}

\usepackage[frozencache=true,cachedir=./_minted]{minted}
\usepackage{xcolor}
\setminted{
  fontsize=\small,       % Smaller font size
  linenos,               % Show line numbers
  breaklines,            % Automatically break long lines
  breakanywhere=true,    % Break long words if needed
  autogobble,            % Remove common leading whitespace
  numbersep=5pt,         % Space between line numbers and code
  frame=lines,           % Add top/bottom frame lines
  framesep=4mm,          % Padding between code and frame edge
  bgcolor=gray!5,         % Light gray background color
  escapeinside=||,
}

\usepackage{multirow}

\usepackage{graphicx} % Required for inserting images
\graphicspath{{paper/}}

\usepackage{natbib} % For bibliography
\copyrightyear{2025}
\acmYear{2025}
\acmConference[UIST '25]{The 38th Annual ACM Symposium on User Interface Software and Technology}{September 28-October 1, 2025}{Busan, Republic of Korea}
\acmBooktitle{The 38th Annual ACM Symposium on User Interface Software and Technology (UIST '25), September 28-October 1, 2025, Busan, Republic of Korea}\acmDOI{10.1145/3746059.3747627}
\acmISBN{979-8-4007-2037-6/2025/09}

\begin{document}

\title{Graffiti: Enabling an Ecosystem of Personalized and Interoperable Social Applications}

\author{Theia Henderson}
\affiliation{
  \institution{MIT CSAIL}
  \city{Cambridge}
  \state{MA}
  \country{USA}
}
\email{theia@mit.edu}
\author{David R. Karger}
\affiliation{
  \institution{MIT CSAIL}
  \city{Cambridge}
  \state{MA}
  \country{USA}
}
\email{karger@mit.edu}
\author{David D. Clark}
\affiliation{
  \institution{MIT CSAIL}
  \city{Cambridge}
  \state{MA}
  \country{USA}
}
\email{ddc@csail.mit.edu}

\begin{abstract}
% Add more "color" to the problem - Josh

Most social applications, from Twitter to Wikipedia,
have rigid one-size-fits-all designs, but building new social applications
is both technically challenging and results in
applications that are siloed away from existing communities.
We present \emph{Graffiti}, a system that can be used
to build a wide variety of personalized social applications
with relative ease that also interoperate with each other. People can freely move between
a plurality of designs---each with its own aesthetic, feature set,
and moderation---all without losing their friends or data.

Our concept of \emph{total reification} makes it possible
for seemingly contradictory designs, including conflicting
moderation rules, to interoperate.
Conversely, our concept of \emph{channels}
prevents interoperation from occurring by accident, avoiding context collapse.

Graffiti applications interact through a minimal \emph{client-side API},
which we show admits at least two \emph{decentralized} implementations.
Above the API, we built a Vue plugin, which we use to
develop applications similar to Twitter, Messenger, and Wikipedia
using only client-side code.
Our case studies explore how these and other novel applications interoperate,
as well as the broader ecosystem that Graffiti enables.

\vfill
\end{abstract}



\maketitle
\newpage
\section{Introduction}

\begin{figure*}[h]
    \centering
    \includegraphics[width=\textwidth]{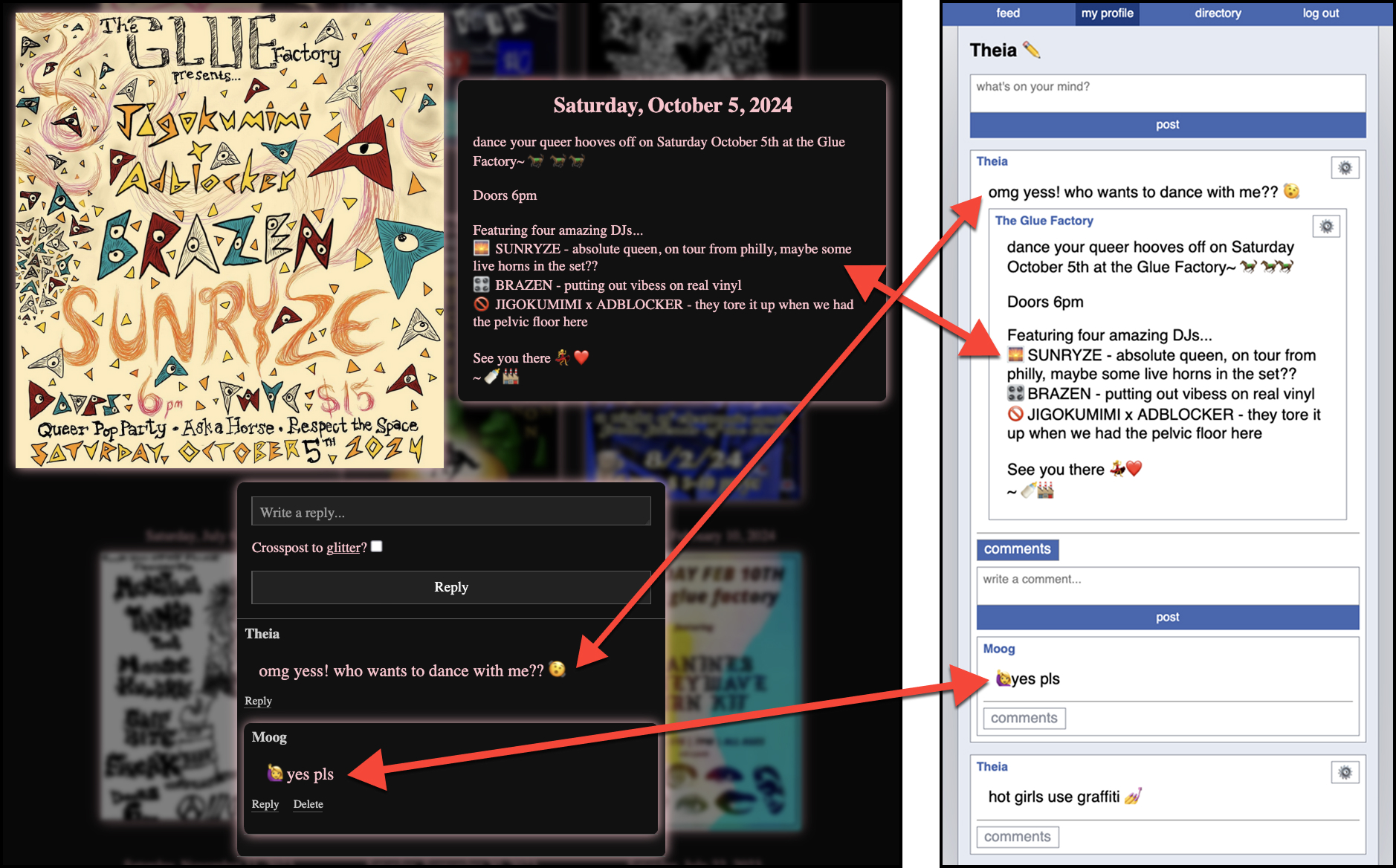}
    \Description{
     The figure shows two social applications with lines connecting components in each of them. The application on the left has an unusual dark neon design and displays an event flyer as well as an event description and replies to the event. The application on the right looks like a more typical micro-blogging application. A reply to the event on the left interface is connected by an arrow to an identical post on the right interface. The post on the right contains a Quote-Tweet style sub-post that is connected by an arrow and identical to the event description on the left. A reply to the reply on the left is connected by an arrow and identical to a reply to the post on the right.
    }
    \caption{
    Interoperation between two applications built with Graffiti:
    The Glue Factory, an application operated by a venue to advertise their shows,
    and Glitter, a microblogging site.
    The show flyer visible in The Glue Factory is used with permission by the artist, Megan Levin.
    }
    \label{case-studies:fig:gloof-and-glitter}
\end{figure*}

Most social applications today are built like international airports,
designed to serve everyone, and soulless as a result.
Where are the online spaces that feel like
the unofficial skate park behind the grocery store,
the dance floor of a long-running nightclub,
or that one friend's living room?
Where are those intentional community spaces that
are not meant for everyone,
but are \emph{everything} to the people
who collectively curate their aesthetic and social norms?

Their absence is partly because truly personalized social applications
%DK given the framing you are using here, I wonder if "social web sites" is better than "social applications" (or if you need to use both and explain why).   Your examples---the nightclib, the friend's living room (and also Gloof) feel like sites not apps.   Per my email, it may be useful or necessary to talk somewhere in the paper about the difference between sites (you go there to interact with a certain kind of contet pertinent to the site?) ad apps (you pull in content from all over to interact with?) and how with gradfiti there is't really a differece
are extremely difficult to build,
and partly because the great power of social media---its
ability to connect massive numbers of people---also
creates social pressure for everyone to use the same few applications.

We present a system called \emph{Graffiti} that makes
building \emph{personalized} social applications
%and social sites with distinctive personalizities
more accessible,
while maintaining the interconnectedness that is special to social media.
We lay the foundation for developers to create a diverse ecosystem
of social applications
%and sites
with only client-side code and with minimal friction
to introducing new features,
while also ensuring that these applications \emph{interoperate},
allowing people to freely migrate between them without
losing their friends or data.
The types of social applications that can be built on the same set of Graffiti primitives range from analogs of Twitter to Facebook to Messenger to Slack to Goodreads to Pinterest to Wikipedia,
with plenty of room for entirely new creations.
%including social ``sites'' with distinctive personalities.

Our goal of interoperability leads almost immediately
to the conclusion that a user's
social data---their posts, likes, friend lists, and so on---\emph{cannot}
be stored within any single application.
Instead, user data is stored on a collective infrastructure
that all applications can access (subject to access control)
and selectively present data from according to their individual designs.

This application-infrastructure separation
is an interoperability mechanism similar to
those used by
ActivityPub~{\cite{activitypub}},
which underlies Mastodon,
and the AT Protocol~{\cite{bluesky}}, which underlies Bluesky.
However, these social protocols---primarily designed to support
Twitter-like networked interactions---are limited in the range of applications
and features they can support.
Some limitations stem from hardwired assumptions,
like per-``instance'' moderation in ActivityPub
or the public-only visibility of content in the AT Protocol.
Other capabilities are technically \emph{possible}
but require standing up custom servers or navigating
a slow and complex standardization process.
% to perform out-of-band interactions.
In contrast, Graffiti provides a simple yet minimally constraining
client-side \emph{application programming interface} (API)
upon which a much broader range of personalized applications can be built, all
without server code or bureaucracy.

A key challenge of Graffiti's design is to resolve
the \emph{tension} between personalization and interoperability.
Allowing applications to have different styling is
not a problem, but can an application with top-down
moderation interoperate with a democratically moderated one?
Is it possible for a small community forum to exist and
not get flooded by the rest of the ecosystem's traffic?
We resolve these questions respectively with two concepts: \emph{total reification},
where all social interactions
are first-class objects that
can be interpreted or ignored by different applications;
and \emph{channels}, which organize data contextually,
preventing a phenomenon called ``context collapse''~\cite{contextcollapse}.

Our design decisions are guided by a set of \emph{requirements}, outlined
in Section~\ref{requirements}, that primarily focus on the direct experience
of social application developers and the users of their applications,
both users of Graffiti as a whole.
We include one additional requirement that targets the indirect
effects that \emph{power} over the underlying infrastructure
can have on the user experience.
Specifically, we leave room for users to retain control over their data
via technologies like decentralization or encryption.

Importantly, however, Graffiti is not shaped by a particular
means of achieving this distribution of power. This again contrasts Graffiti
with systems like ActivityPub and the AT Protocol,
which are inseparable from their decentralized architectures
and leak architectural details, like ``instances'' and ``relays,''
into their APIs.
Graffiti's API-first\footnote{
We reply to the provocation ``\emph{Protocols, Not Platforms}''~{\cite{protocolsnotplatforms}},
the rallying cry that sparked Bluesky~{\cite{bluesky_from_protocols}},
with the amendment, ``\emph{...but the API first!}''
} approach not only reduces complexity for
developers who simply want to make applications,
but also allows the infrastructure underneath
to update as both technologies and threats evolve.
In fact, the Graffiti API is designed so that multiple implementations can exist
\emph{in parallel} underneath it,
allowing for the incremental adoption of new technologies,
similar to the web's transition from HTTP to HTTPS.

After our requirements, we describe the Graffiti API,
first through high level concepts in Section~\ref{concepts},
then in detail in Section~\ref{api}.
Section~\ref{above-and-below} describes
the decentralized protocols we have built \emph{below} the API,
as well as the tooling we have developed \emph{above} the API
to build Graffiti applications with declarative and reactive programming.
In Section~\ref{case-studies},
we evaluate Graffiti through a series of case studies, demonstrating
how it is possible to build applications like Twitter, Messenger, and
Wikipedia, as well as novel applications on Graffiti, and how unusual
variants of all of these interoperate.
Finally, we wait until Section~\ref{related-work}
to continue exploring related work, by then with a more concrete
understanding of what we are relating to.

\newtheorem{requirement}{Requirement}

\section{Design Requirements}
\label{requirements}

Graffiti's design is primarily driven by our
first two requirements:
(\ref{requirements:personalization})
the need for communities to personalize
their social applications,
and (\ref{requirements:adversarial-interop})
the need for those applications to interoperate.
Most other requirements work to clarify these two.

Requirements \ref{requirements:autonomous-extensibility}
and \ref{requirements:serverless} expand on how accessible
personalization should be: it should be possible to add new features
without a lengthy standards process, and possible to build
applications without writing server code.
Requirements \ref{requirements:context-differentiation}
and \ref{requirements:forgiving}
address possible antisocial consequences of
interoperation that we need to avoid,
namely by protecting users from ``context collapse,''
and ensuring their ``right to be forgotten.''

Requirement~{\ref{requirements:parallel-implementations}} stands apart to address
how Graffiti might be implemented to uphold these requirements over time, despite
the natural tendency for social technologies to ``enshittify''~{\cite{enshittification}}.

\begin{requirement}[Personalization]
\label{requirements:personalization}
   It should be feasible for communities to build
   their own social media applications.
\end{requirement}

Over the past several years,
works including
The Three-Legged Stool~\cite{threeleggedstool},
runyourown.social~\cite{runyourownsocial},
and pluriverse.world\footnote{\url{https://pluriverse.world}}
have opined the importance of relatively small,
community-specific social applications.
Existing examples of such applications include the fanfiction
repository \href{https://archiveofourown.org/}{Archive of Our Own},
the hypermedia curation platform \href{https://www.are.na}{are.na},
the movie review site \href{https://letterboxd.com/}{Letterboxd},
and Vermont's local \href{https://frontporchforum.com/}{Front Porch Forum}~\cite{threeleggedstool}.

From their styling, to the features they include (and the ones they omit),
to their moderation policies, to their social norms, these applications
are a reflection of the communities they serve and have a
distinct sense of being \emph{lived-in}
that one-size-fits-all platforms lack.
Most importantly, these applications are designed in close cooperation with the communities themselves,
giving members a sense of empowerment and autonomy over their shared space.
Proponents argue that such online communities are the key to a healthier
digital public sphere and the lessons members learn by participating in such
communities have benefits for civic society more broadly~\cite{threeleggedstool, runyourownsocial, archiveoftheirown}.

The issue is that such applications are difficult to build and maintain,
requiring either significant technical expertise or the money to outsource the work.
Unprivileged communities must resort to other means like creating group chats,
subreddits, or Facebook groups. %, or Mastodon instances.
While these services host many thriving communities,
those communities are often faced with rigid and inadequate feature sets,
limited customization options, and shifting platform designs and policies.
We aim to empower \emph{all} communities with the ability to
build their own applications, giving them autonomy over their digital spaces.

Of course, that is a lofty goal. While the system
we later describe lowers the barrier to participation,
it still requires creating or modifying front-end code, although
we outline steps in Section~\ref{above-and-below:above:future-work}
to lower this barrier further.
This requirement and the ones to follow exist on a spectrum.
We attempt to design for all of them, but are limited by
the scope of this work and the tension that the requirements put
on each other.

\begin{requirement}[Adversarial Interoperability]
\label{requirements:adversarial-interop}
Any application \emph{A} can be built to interoperate with an application \emph{B} without any compliance from application \emph{B}.
\end{requirement}

Communities are not monoliths; they are composed of individuals
with different and evolving preferences.
With interoperability, it becomes possible for individual users and groups of users to explore new application designs
without breaking from their existing communities.
This is especially important when, online, a ``small''
community may consist of tens of thousands of people.

For example, it should be possible for a group of users to
modify an application so that it only shows five posts per day (to combat screen addiction),
without imposing this feature on all other users of the original application.
The original application may eventually include this feature in a configuration panel,
or the applications may drift further apart as their users' needs diverge. However, this
evolution should be driven entirely by users rather than by external pressures
to either conform to a particular design or abandon community altogether.

We use the term ``adversarial interoperability''~\cite{adversarialinterop} because it may not be in the
interest of a large existing application to interoperate with a new application that may co-opt some of its market share.
For example, Facebook, X, and Reddit all have a history of intentionally breaking
the interoperability of third-party applications by changing or restricting access to their APIs.
Even in small group chats, where the power dynamics are less extreme,
an adminstrator-administee hierarchy
can embody an ``implicit feudalism'' that some claim can even
normalize monarchical power structures in society at large~\cite{governablespaces}.
It should never be possible for one entity to hold users
and their data hostage to a particular experience.
Empowerment should not be limited to a community ``as a whole'' but
distributed amongst its subgroups and individual members.

\begin{requirement}[Autonomous Extensibility]
\label{requirements:autonomous-extensibility}
    An application should be able to freely introduce new social features without requiring compliance from existing applications or the system as a whole.
\end{requirement}

%DK it isn't entirely clear how this requirement differs from the "adversarial interop" one or how "expanding to reactions" differs from "limiting to 5 posts per day".
In classic internet forums, users might reply to a post with ``+1'' to signal their support.
In 2009, Facebook created the like button and in 2016 expanded it to a set of five reactions.
Messaging clients, like Signal, now allow users to react to a message with any single emoji and
Slack allows for users to react with multiple emojis, including ones with custom animated icons.

Social applications evolve in unpredictable ways,
so to avoid instant obsolescence, Graffiti must be highly extensible,
allowing for new features
to be developed with relative ease.
Importantly, it should be possible for one application to introduce a new feature
without requiring compliance from other interoperating applications, to avoid
lock applications into a particular set of features,
just as an ecosystem without interoperability can lock users into a
particular set of applications.

For example, if a developer wants to introduce the feature
``reactions to reactions,'' they should not need to draft an RFC,
convince a group of server admins to implement the feature\footnote{
    See Section~\ref{related-work:activitypub}
},
and rewrite the reaction ontology~\cite{ecosystemmoving, herdingdnscamel, semanticwebtwodecades}.
They should be able to build the feature into their application
entirely \emph{on their own}, with other applications able to adopt the
feature as they see fit.

\begin{requirement}[Serverless]
\label{requirements:serverless}
    It should be possible to build a new social application without writing or deploying any server code.
\end{requirement}

Arguably the most difficult part about writing a novel
social application is writing, deploying,
and maintaining server code.
Signal's cofounder wrote
``people don’t want to run their own servers, and never will''~\cite{moxieweb3}.

Of course, Graffiti must involve servers \emph{somewhere} to store data
and pass it around, but we require that all such servers are \emph{generic}
and not tied to any one particular application.
For example, an RSS reader is a ``serverless'' application since
no new server needs to be built specifically for it to function.
The client-side reader simply polls existing RSS servers, each of which outputs data in a generic format.
In Graffiti, both reading \emph{and writing} content must be done serverlessly.

%DK suggested change:
Certain applications have compute or storage demands that make this requirement impossible to meet.
For example, constructing something like TikTok's For You feed requires
analyzing billions of posts, functionality that is not built into our infrastructure and cannot be implemented by client-side applications.
This work focuses on the broad range of applications
that do not require massive machine learning algorithms,
but we conclude with a brief discussion of extending Graffiti to enable them, too.

\begin{requirement}
\label{requirements:context-differentiation}
    Users should be protected from context collapse.
\end{requirement}

Context collapse is a phenomenon that describes the negative consequences of flattening
a person's multiple audiences into a single context.
Context collapse is a well known problem of existing social applications.
For example, if a person's family holds different political beliefs than their friends
but both groups follow them on Facebook, that person may choose to self-suppress their political
speech to avoid angering either side of their collapsed audience.
Either self-suppressing to appeal to the ``lowest-common denominator'' or ignoring the collapse
and simply posting to an ``imagined audience'' can lead to harm for both
the users involved and society at large~\cite{contextcollapse, contextcollapseimpact, contextcollapsequeer, spiralofsilencesocialmedia}.

Users currently leverage the ``siloing'' of existing applications to manage contexts---using
one application to stay in contact with family and another to connect with friends~\cite{whatsappforfamily}.
Interoperability between applications introduces the possibility
of collapsing all of these silos.
For example, both Tinder, a dating app, and Linkedin, a professional networking app, involve the authoring
and browsing of user profiles.
These user profiles have many semantic similarities (name, biography, employment status) and so
it would not be unreasonable to imagine that in a system of interoperable applications,
the profile a user created in one application
naturally shows up in the other.
Of course, this could be confusing or highly embarrassing, and might even get someone fired.

Graffiti must give users affordances to mitigate traditional sources
of context collapse (\emph{e.g.} the ability to limit their posting to suitable subreddit-like scopes)
as well as affordances to mitigate new sources of inter-application context collapse.

\begin{requirement}[Forgiving]
\label{requirements:forgiving}
    Users should be able to delete, edit, and repudiate content they post.
\end{requirement}

Shortly before Elon Musk initiated his acquisition of Twitter, now X,
he created a poll that asked users whether Twitter should add an edit button.
Out of over four million respondents, 73.6\% voted ``yse'' [sic] and 26.4\% ``no.''\footnote{
  X now lets users edit their tweets within 30 minutes of posting but only
  if they pay for X Premium.
} These results are not surprising as many users of Twitter and Facebook regret
their posts for a variety of reasons
including misjudging social norms, posting while extremely emotional or inebriated,
or posting to an unintended audience (cf. context collapse)~\cite{regret, regrettwitter}.

The right to delete one's data was enshrined into law by the European Union's
General Data Protection Regulation as
the \emph{right to erasure}, previously called the \emph{right to be forgotten}~\cite{gdpr}.
We require a stronger claim, that users should also be able to both \emph{edit}
and \emph{repudiate} content that they have posted. Repudiation is the ability to \emph{disown},
as one might do when they say ``that screenshot has been photoshopped,''
or more recently ``that's AI.'' Repudiation is usually possible in centralized platforms
but can be limited by technologies employed in some decentralized platforms,
like public key signatures~\cite{offtherecord}.

The fact that we require mutability (editing and deleting) and repudiation does not rule
out the possibility that users can \emph{opt in} to services that make their content more permanent.
Technologies exist to make a ``forgiving'' system unforgiving but not the other way around.
For example:
\begin{enumerate}
\item
A post could link to content stored on an immutable storage service like IPFS
or a blockchain to prevent erasure.
\item
When a user replies to a post, their application could include the hash
of that post in their reply. Compatible applications would
hide or put a warning over replies to posts whose hashes do not match,
indicating that the post has been edited.
\item
A user could choose to sign their messages with a PGP signature,
as some people do with their emails, to make them non-repudiable.
\end{enumerate}
Some users, like politicians, may face social pressure to use applications
that employ some of these technologies as a form of accountability.
However, to leave the most room for personalization, forgiveness should be the default.

\begin{requirement}[Parallel Implementations]
\label{requirements:parallel-implementations}
    Applications should be able to send and receive data
    from multiple independent implementations of Graffiti.
\end{requirement}

If a single organization were in charge of hosting Graffiti,
history has shown that, even if it started out by making promises of
personalization and interoperability,
over time it would ``enshittify''~{\cite{enshittification}}
and its promises would erode in favor of profit.
This enshittifaction could also extend to other harms,
such as surveillance.
Therefore, we want to ensure that Graffiti can be implemented using
technologies like decentralization and end-to-end encryption
to protect users and their data.

However, just as we do not want to lock users into a particular application (Requirement~{\ref{requirements:adversarial-interop}}),
or developers into a particular set of features (Requirement~{\ref{requirements:autonomous-extensibility}}),
we do not want to lock the system itself into a particular implementation.
The ``ecosystem is moving''~{\cite{ecosystemmoving}}, with both rapidly evolving
technologies and threats, and
leaning too heavily into any one solution would set Graffiti up for failure.

Instead, we require that Graffiti supports multiple implementations
\emph{in parallel}.
This is similar to the World Wide Web, which was designed according to the
``principle of minimal constraint'' to allow for competing implementations,
such as FTP, to coexist with HTTP~{\cite{weavingtheweb}}.
This is made possible by the fact that URLs begin with a \emph{scheme},
such as \texttt{http:}, indicating which implementation should be
used to retrieve the resource.
This allows new implementations, such as HTTPS, to be
incrementally adopted without invalidating existing ones.

\section{Design Concepts}
\label{concepts}

This section outlines the primary concepts that make up the design of Graffiti
according to our requirements.
We follow with Section~\ref{api}, which outlines the \emph{specific} API that is built out of these
concepts.

At a high level, the Graffiti API works as follows.
\emph{Actors} (basically users) create \emph{objects} which can represent
social artifacts like posts or \emph{activities} like ``likes.''
Objects are posted to a set of \emph{channels} and other users can
discover an object by querying one of its channels. Channels are a flexible representation
of different \emph{contexts} which may include identities, topics, locations,
and more.
Actors cannot modify other actors' objects and so for actors to interact with each others' content,
those interactions must be \emph{reified} into individual
activities,
allowing for different interpretations of them.
Objects can be configured to be either public
(to anyone observing its channels),
or private to a set of users on its \emph{allowed list},
an additional affordance for context management.

\subsection{Objects}
\label{concepts:objects}

Objects are the atomic units that encapsulate \emph{all}
of the social data in Graffiti, including
both social artifacts (such as posts and profiles) and social ``activities'' (such as likes and follows).
Each Graffiti object includes some structured metadata,
such as the actor who created the object,
but an object's ``value'' can be \emph{any} valid JSON object~\cite{json}.
For example, a blog post object may have the value:

\begin{minted}{javascript}
{
  title: "My First Post",
  content: "Hello, world!",
  published: 1611972000, // The time in seconds
}
\end{minted}

This idea of encapsulating social data in JSON objects is
taken directly from the ActivityStreams standard~\cite{activitystreams},
and many common \emph{properties} for objects,
including the ones above, are defined in the
ActivityVocabulary~\cite{activityvocab}.
However, we choose not to inherit the ``linked data'' piece
of these specifications which makes it complex to create new object properties.
Instead, if a developer wants to make a recipes app, for example,
they can freely add an \texttt{ingredients} property to their objects.

We are leveraging an idea here called a \emph{folksonomy}~\cite{folksonomy}.
Social applications
benefit from network effects, and so we expect that developers will make use of
existing properties where possible,
but will introduce new properties when necessary.
The result will inevitably be messy and inconsistent, but, as with the evolution of
particular hashtags, we expect there will be an effective balance of creativity and convergence.
Importantly, the barrier to participating in the folksonomy is low
and does not require a precise ontological agreement,
which some argue contributed to the failure of the semantic
web~\cite{semanticwebtwodecades}.
The folksonomy approach encourages autonomous extensibility,
per Requirement~\ref{requirements:autonomous-extensibility}.

Graffiti objects also have URLs that
start with a \emph{scheme}, just like URLs on the web,
enabling objects to be served by different \emph{parallel} implementations,
per Requirement~{\ref{requirements:parallel-implementations}}.
For example, a post might have the URL
\texttt{graffiti:\allowbreak{}remote:\allowbreak{}pod.\allowbreak{}graffiti.\allowbreak{}garden/\allowbreak{}123},
where the scheme \texttt{graffiti:\allowbreak{}remote:} refers to an implementation that we describe
in Section~\ref{above-and-below:remote-protocol}
instead of \texttt{https:} as seen on the web.
URLs allow the object to be directly fetched%
% \footnote{
%   In Section~\ref{above-and-below} we describe how fixed URLs
%   are still compatible with ``data portability,'' allowing
%   a user to move their data to a new place or protocol without
%   changing the URL.
% }
, but also pointed to, as in the
``like'' activity value below.

%DK ATP: there's a trick I think medium uses that we might want to steal.   they create urls from the title of a post, but *also* include in the url a weired number that I assume is a nonce.   If you change the title of your post, it changes the title part of the url but not the nonce.   Which means that you can look up the post using either the old or the new url and both will work because the medium server is smart enough to look only at the nonce to decide what content is wanted.   Could we do something similar, so that if the user later decided to change the delivery protocol it would possible to interpret old urls (possibly with some help from the user's profile) in a way that would usually still let you fetch the contents?   For example, it would be nice of the user's profile could store a "redirect rule" explaining how to generate the current url from the old

\begin{minted}{javascript}
{
  activity: "Like",
  target: "graffiti:remote:pod.graffiti.garden/123",
}
\end{minted}

This like activity is an example of \emph{reification},
a technique frequently used in the semantic web community~\cite{rdfprimer}.   We use reification to represent actions on data as data itself.
Rather than a post object having a ``like'' method that mutates metadata about the post,
the ``like'' here is a separate object that points to the post object.
Reification allows for new interaction mechanisms to be introduced as easily as new properties,
without changing the underlying system.
While AcitivityStreams and other systems use reification,
Graffiti employs \emph{total reification} where
all interactions, including moderation actions, are reified,
which we discuss in
Section~\ref{concepts:total-reification}.

Finally, objects are mutable, meaning they can be changed or deleted.
Mutability is necessary for the system to be forgiving, per Requirement~\ref{requirements:forgiving}.
The question of \emph{who} can mutate the objects
will be answered once we discuss identity, next, and then total reification.

In summary, Graffiti objects are:

\begin{enumerate}
\item
Flexible: Objects can take on any value according to a self-describing folksonomy.
\item
Globally-Addressable: Every object has a globally unique URL that can be used to fetch or point to it.
\item
Mutable: Objects can be changed or deleted.
\end{enumerate}

\subsection{Actors}
\label{concepts:actors}

An actor is an account that a user can ``log in'' as
to identify themselves, publish objects, and access private objects,
so long as they are on that private object's ``allowed'' list
(to be discussed in Section~{\ref{concepts:allowed-lists}}).
The actor is represented by a globally unique string or \emph{URI},
just as an email account is represented by an email address.
The term \emph{actor} is again inherited from ActivityStreams,
and is used rather than \emph{user}, since a user may own
multiple actors or
an actor may be a non-human entity
like a bot or recommendation service.
All Graffiti objects are associated with an actor, the actor
that created the object, but anonymous
or meronymous~\cite{meronymous} interactions are possible by
creating ``throw away'' actors.

Unlike an ActivityStreams actor, a Graffiti actor does not embody
%DK there is no information that must be associated with an actor other than...
any identifiable information other than its
URI.
Properties like a display name or pronouns are instead
defined in separate objects.
For example, if Alice's actor URI is
\texttt{https://\allowbreak{}alice.\allowbreak{}example.\allowbreak{}com}, a profile object value for her is:
\begin{minted}{javascript}
{
  name: "Alice",
  pronouns: "she/her",
  describes: "https://alice.example.com",
}
\end{minted}
%DK how do you prove Alice owns the actorId; prevent someone else from pretending to?  Does the actorid come with a key alice can use to sign her profile object?

This separation allows for new profile
properties to freely evolve
and it also allows people to have different profiles for different contexts.
For example, a trans person who is not publicly ``out'' may want to
change their name within their queer community without
changing their name to the broader public.
Alternatively, the separation allows users to create nicknames
or ``petnames'' for each other~\cite{petnames}.

%DK there's some inaccurate implication here: while channels can provide unlisted groups (soft access control) there is no mechanism for specifying that access is only *permitted* (hard access control) by actors that are defined to be members of a particular group.
%ATP Or maybe I should call it an asymmetry?   If we really committed to capability based access control, we wouldn't use hard access control by actor id.   instead we would allow an actor to craft a "key" an specify that certain objects are unlocked by that "key" and anyone with the key has access to those objects including to mutate them?

\subsection{Total Reification}
\label{concepts:total-reification}

Within Graffiti, not some but \emph{all} interactions between actors
are \emph{reified}.
This \emph{total reification} allows different applications to have
different ``rules'' regarding what users are ``permitted'' to do, while
still interoperating.

For example, suppose a developer would like to create an application
that interoperates with an existing application,
but that allows certain actors (moderators) to remove other actors' posts.
If the moderators' removals are reified into \texttt{"Remove"} activities,
then the original application can simply ignore the activities
while the new application can attend to the new signal and hide ``removed'' posts from the display.
To be clear: an offending object is not removed from Graffiti as a whole,
it is just removed from any application that understands \emph{and respects}
the moderators' \texttt{"Remove"} activities.%

These moderation rules can evolve as developers
introduce new objects and properties,
or new procedures to interpret those objects.
For example,
perhaps the set of moderators is elected by actors who vote with reified \texttt{"Vote"} activities.
Or, perhaps an application hides posts that exceed a certain threshold of \texttt{"Remove"} activities,
regardless of whether the removing actors are considered moderators.
Applications may also consider authors of posts to be the \emph{de facto} moderators of the replies to their
own posts or allow them to disable replies altogether by adding a \texttt{disableReplies}
property to their post objects.
A fully democratic system that allows actors to introduce and ratify new moderation proposals,
as described by the PolicyKit project~\cite{policykit}, could also be reified into
individual-actor activities.
Importantly, all of these systems can coexist in different interoperating
applications and users can freely opt in or out of different systems simply by switching
the application they use.

The flexible object values that we described in Section~\ref{concepts:objects} make arbitrary reified interactions \emph{possible}, but we use the term \emph{total} reification because in Graffiti this is the \emph{only} way to represent interactions.
We impose this rule to achieve the \emph{adversarial} part of our adversarial interoperability requirement (\ref{requirements:adversarial-interop}).
If an application could modify the \emph{actual} state of an object, then it could force all other applications to accept its decisions about the object.
For example, if an actor could actually \emph{delete} another actor's post
from Graffiti \emph{as a whole}
then it would become impossible to build
an application that does \emph{not} recognize the deleting actor as a moderator.
Thus, we ensure that:
\begin{enumerate}
\item
Objects are \emph{proprietary}: Only the creator of an object can modify or delete it.
\end{enumerate}
Conversely, adversarial interoperability means that no applications should be able to ``block'' another from applying an interaction, which means that it must be possible for an actor to publish
their reified action object somewhere that is visible
to relevant applications.
Graffiti objects are published in channels, described in the next
section, and therefore:
\begin{enumerate}
\setcounter{enumi}{1}
\item
Channels are \emph{permissionless}: actors can freely read from and write to channels that they know of.
\end{enumerate}

\subsubsection{Antipaternalism}

Total reification grants people the ability to keep content off of ``their'' applications,
but it prevents anyone from keeping content off of Graffiti as a whole\footnote{%
It \emph{is} possible for content to be deleted in implementations ``below'' the API,
allowing hosting providers to delete illegal content,
like Child Sexual Abuse Material (CSAM).
However, if the implementation is end-to-end encrypted,
it may not be possible for the hosting provider to
detect illegal content,
a fundamental dilemma of encryption technology.
}.
While this antipaternal stance may seem extreme,
it is no different than the web or other open infrastructures,
like ActivityPub or the AT Protocol:
people can always find \emph{some}
web server, ``instance,''
or ``personal data store'' to host their content.

However, our design is more than a reluctant acceptance of the way things are.
In line with harm reduction practices~\expandafter{\expandafter\cite{harmreduction}},
there are benefits to keeping harmful content---which will inevitably
exist---within Graffiti as a whole, while filtered from mainstream spaces.
For example, when users drawn into extremist ideologies are not
siloed away on entirely disconnected systems
like 8kun or Gab~\cite{8kun,gab}, it becomes possible to create off-ramps
to help them disengage gradually without going ``cold turkey.''
Of course, further work is needed to assess the efficacy of
such a strategy and its possible harms.

\subsubsection{Accountability}

Total reification works to enforce ``adversarial interoperability''
per Requirement~{\ref{requirements:adversarial-interop}}
and thus to protect users from leaders---even of small communities---that abuse their role as
``benevolent dictator for life''~{\cite{governablespaces}}.
Our solution matches the suggestion given by Melder et al.
in their study of Fediverse communities with imperfect leadership
to introduce a ``user-level selective ability to bypass [moderator] filters''~{\cite{blocklistboundary}}.
However, it is worth considering whether total reification makes it so easy
to opt out of a community's norms that users are not held accountable for their actions.

While there is work to be done studying this in the context of Graffiti specifically,
work by Hessel et al.
suggests the opposite effect~{\cite{highlyrelatedcommunities}}.
They find that users who explore parallel community spaces
with different norms not only still meaningfully participate in their original community space,
but also become more engaged in it.

\subsection{Channels}
\label{concepts:channels}

Channels are the mechanism Graffiti uses to notify
actors of new objects.
Rather than sending objects directly to specific actors\footnote{
    See our discussion of ActivityPub's ``Actor Model'' in Section~\ref{related-work:activitypub}.
},
channels are a publish-subscribe mechanism~\cite{pubsub}
where objects published to a channel
can be read by actors subscribing to that channel.
However, unlike most publish-subscribe systems,
an object remains ``in'' a channel until
it is explicitly removed from that channel
or the object as a whole is deleted.
Therefore, we can consider a channel both a ``place''
an object can be and a property of the object itself.

The channel namespace is \emph{global} with every string (up to a certain length)
naming exactly one channel.
Like object properties, these channel names have no inherent meaning
other than the meaning ascribed to them by developers or users.
In practice, the meaning of a channel is a particular social \emph{context} and the channel
name is some sort of \emph{shared information} implicit to that context.
For example, the channel named by an object's URL is a natural place to post content ``\emph{talking about}'' that object, like replies, and we expect that most applications will follow that norm.
Importantly, content within channels is effectively hidden from applications
that do not know their names,
an affordance to prevent context collapse, per Requirement~\ref{requirements:context-differentiation}.

We will first illustrate the power of channels
to model different types of \emph{replies}.
Then, we describe the formal properties of channels and their other uses.

\subsubsection{Replies}

Most social applications allow users to reply to posts, but there are surprisingly subtle
contextual differences between different reply designs.
For example, on Instagram, replies to a post are only ever displayed to people viewing that particular post\footnote{
Instagram used to have a ``Following'' tab where a user could see the replies that people they follow had posted across the application.
This led to numerous examples of context collapse and so Instagram discontinued it in 2019~\cite{instagramfollowingtab}.
}, while on X, a user's replies from across the application can all be found in their Replies tab.
These differences reflect the platforms’ contrasting design goals of intimacy versus reach.

With channels, \emph{both} designs can \emph{co}exist:
\begin{itemize}
\item
To create an Instagram-like reply,
an application should post the reply to the channel
named by the original post's URL---for brevity, the ``post's channel.''
Applications displaying the original post implicitly know that post's channel name
and can query the channel to populate the reply thread.
\item
To create an X-like
reply, an application should post the
reply to \emph{both} the post's channel
and the channel named by the replier's actor URI---for brevity, the ``replier's channel.''
Applications displaying the replier's profile implicitly know that replier's
channel name and can query the channel to populate the Replies tab.
\end{itemize}

Finally, what happens if the reply is \emph{only} posted to the
replier's channel?
In that case, the reply becomes like a Quote Tweet\footnote{
Prior to 2020. \url{https://x.com/Support/status/1300555325750292480}
}: the reply is visible to the replier's followers but
not the audience of the original post, as they do not have implicit
access to the replier's channel.
A summary of these different designs is shown in Table~\ref{concepts:channel-replies}.

\begin{table}[htbp]
\centering
\caption{Reply designs according to channel usage}
\label{concepts:channel-replies}
\begin{tabular}{cc|c|c}
\multicolumn{4}{l}{\emph{Is the reply posted to the...}} \\
& \multicolumn{1}{c}{} & \multicolumn{2}{c}{\emph{\textbf{...post's channel?}}} \\
& & {\checkmark} & {$\times$} \\
\cline{2-4}
\multirow{2}{*}{\shortstack{\emph{\textbf{...replier's}} \\ \emph{\textbf{channel?}}}}
& {\checkmark} & X Reply & Quote Tweet \\
\cline{2-4}
& {$\times$} & Instagram Reply & N/A
\end{tabular}
\end{table}

Importantly, these reply designs interoperate in a way that allows content producers to target a particular audience,
regardless of the applications others choose to use:
Instagram-like replies will not be displayed in X-like Replies tabs.

Therefore, channels can be considered a constraint on application personalization,
in tension with our overarching Requirement~\ref{requirements:personalization}.
We introduce this constraint because, unlike total reification---where applications
can present different interpretations of the underlying
content simultaneously, e.g. a post is ``removed'' in one application but not in another---a reply
cannot be simultaneously ``seen'' and ``not seen'' by a potential audience member.
We prioritize giving posters the ability to express their ``targeted imagined audiences''~\cite{imaginedaudience}
to mitigate context collapse, at the cost of slightly reducing consumer autonomy.

\subsubsection{Properties}

In summary, channels are:

\begin{enumerate}
\item
Persistent: When an actor publishes an object to a channel, other actors can
continue to retrieve the object until it is removed from the channel or deleted.
\item
Global: The channel namespace is global with every string (up to a certain length)
corresponding to exactly one channel.
\item
Crosspostable: An object can be associated with more than one channel (or no channel at all).
\item
Obscure: It is not possible to read from or write to a channel without knowing
its name.
\item
Permissionless: Actors can freely read from and write to any channel \emph{that they know about}.
\end{enumerate}

Our emphasis on the last property, which we introduced in our discussion of
total reification, is due to the additional ``Obscure'' property
which allows channels to provide a form of \emph{security by obscurity}, like an ``unlisted'' link.
However, security by obscurity does not allow for \emph{revocation}\footnote{
    See Section~\ref{related-work:ocaps} for a discussion of how channels relate to ``object capability security''
    which is similar to security by obscurity but allows for revocation.
}
of access,
which is why objects may also be access-controlled,
as we will discuss in Section~{\ref{concepts:allowed-lists}}.

Despite the fact that the ``Obscure'' property implies that
there is no global directory of channels,
many channels may be found by spidering Graffiti like a search engine.
Therefore, a determined
developer could make an application that intentionally collapses contexts.
This is an unfortunate possibility with any public content.
For example, snoopreport.com sells a context-collapsing service for Instagram.

Given this, users with privacy critical needs should, again, use the
access control affordances described in Section~{\ref{concepts:allowed-lists}}.
For less critical needs,
we hope to cultivate social norms that dissuade scraping, similar to informal notions of privacy in the real world.
For example, it is entirely possible for a person in the real world
to attend various support groups, like alcoholics anonymous,
just to spread people's secrets, but doing so would be shameful.
On the web, \texttt{robots.txt} is a functional socially-enforced standard
to prevent web scraping.

\subsubsection{Other Channel Meanings}

We have seen how channels can represent \emph{people} and \emph{posts}
which cover the contextual needs of most microblogging applications,
but channels can also represent media, topics, websites, and locations,
and more meanings can be freely introduced in an evolving folksonomy~\cite{folksonomy}.
A table of possible channel meanings and the types of applications
those meanings can enable is shown in Table~\ref{concepts:channels-and-applications}.

\begin{table}[h]
\small
\caption{Applications enabled by channel meanings}
\label{concepts:channels-and-applications}
\centering
\begin{tabular}{|l|l|}
\hline
\textbf{Channel Meanings} & \textbf{Representative Applications} \\ \hline
External Media & Letterboxd, GoodReads, Genius \\ \hline
User Media & YouTube, Medium, Github \\ \hline
Topics & Reddit, Discord, Slack \\ \hline
Collections & are.na, Pinterest, Spotify playlists \\ \hline
Websites & Hypothes.is, Pinboard \\ \hline
Physical Locations & Tinder, Craigslist, Nextdoor \\ \hline
Virtual Locations & Second Life, Minecraft \\ \hline
Events & Partiful, Facebook Events \\ \hline
Products/Services & Amazon reviews, Yelp \\ \hline
Documents & Wikipedia, Google Docs \\ \hline
\end{tabular}
\end{table}

In some cases, an existing standard can be used as a channel
name, like the URL of a website, for Hypothes.is analogs,
or the ISBN of a book, for GoodReads analogs.
In other cases, the channel name may be user generated,
as with public contexts like subreddits or Wikipedia pages,
%DK as discussed, it is problematic to use human-readable names for subreddits, since it means you can never change them.   and while i can see some limited benefit to human-readable wikipedia names, i don't think it extends to subreddits.   Much better to have a directory page of subreddits and their corresponding channels.
or randomly generated for ``unlisted'' contexts like
Discord channels, Pinterest boards, or Google documents.
We discuss how continuous contexts, like geographic locations, can be
mapped onto specific channel names in Appendix~\ref{appendix:more-channel-names}.

\subsection{Allowed Lists}
\label{concepts:allowed-lists}

Allowed lists are an access control mechanism
that an actor can optionally apply to an object
to limit its visibility to a specific list of actors.
Allowed lists enable direct messaging and private groups,
as well as ``visibility control''~\cite{visibilitycontrol, toogayforfacebook}
affordances in networked settings such as
Facebook-like ``Friends''-only posts.

We use this simple whitelist mechanism of access control rather than,
say, Unix file types, because the additional affordances that
such a system would provide can be built more flexibly with total reification.
For example, in Section~{\ref{case-studies:parallax}} we show how mutable group
membership in a messaging application can be constructed.

Allowed lists do not replace channels when they are used
but rather complement them,
with allowed lists providing guaranteed privacy
and channels specifying the context in which the content should appear.
For example, a friend group may communicate via a group chat,
plan a trip via a private spreadsheet, and post ``Close Friend'' stories
for each other to see.
Objects representing interactions in these different scenarios
would have identical allowed lists,
but objects would be placed in different channels based on their context.

\section{API}
\label{api}

We now describe our API that fulfills the requirements
outlined in Section~\ref{requirements} and builds upon the concepts
just discussed in Section~\ref{concepts}.
The API features simple methods for \texttt{login} and \texttt{logout}
and a series of CRUD\footnote{Create, Read, Update, Delete} methods for manipulating objects:
\texttt{put}, \texttt{get}, \texttt{patch}, and \texttt{delete}.
The novel \texttt{discover} method allows users to query for objects
from a set a of channels.
Finally, there are two specialized ``recovery'' methods included for completeness:
\texttt{recoverOrphans} and \texttt{channelStats},
which prevent users from ``losing'' objects or channels, respectively.
Overall, as much complexity as possible is pushed \emph{below} the API,
while as much functionality as possible is pushed \emph{above}.

The API is written as an abstract TypeScript class, allowing
developers building on top of the API to swap out different implementations
with a single line of code.
The API's complete documentation, specification, and test suite is available online\footnote{
    \url{https://api.graffiti.garden/classes/Graffiti.html}
}, and so here we focus on the important design decisions.

TypeScript is a type annotation layer around standard JavaScript
and provides enormous benefit to developers working on large projects.
However, the Graffiti API is still completely functional
in a regular JavaScript environment and entire Graffiti applications can be written
with vanilla HTML, CSS, and JavaScript that function in the browser without
any build step.
These web technologies can also be used to build desktop applications
with Electron\footnote{
    \url{https://www.electronjs.org/}
} or mobile applications as progressive web applications\footnote{
    \url{https://developer.mozilla.org/en-US/docs/Web/Progressive_web_apps}
}.

\subsection{Logging In}

While it is possible to retrieve public content without logging in,
all other functionality, including
creating and modifying content or reading access-controlled content,
requires proof that the user owns a particular \emph{actor}, thus requiring a login.

Modern log in methods, like OAuth 2.0, while secure, are incredibly complex~\cite{oauth}.
To hide this complexity from application developers and to make the API future proof
to evolving security standards, the API provides a simple \texttt{login} method that works as follows:

\begin{enumerate}
\item
A user performs an action, like clicking a ``log in'' button, that triggers
the application to call the \texttt{login()} method.
\item
In response, the implementation of the API may do any manner of things,
including opening up a pop up or redirecting the user to a new page.
In that \allowbreak{implementation-specific} interface, which the application developer
does not need to build, the user might be asked to provide credentials like a username
and password or a public key pair.
See Figure~\ref{above-and-below:figure:login} for our implementation.
\item
Once the user is authenticated they are redirected back to the application, if applicable,
and a JavaScript \emph{event} fires that includes
a \texttt{session} object. The \texttt{session} object includes the authenticated \texttt{actor}
URI and other information that the implementation needs to verify authentication, like
a token or a signing function, however those details are entirely opaque to the application developer.
\item
When the application wants to perform an action that requires authentication,
it must provide the \texttt{session} object. For example,
\texttt{delete(object, session)}.
\item
Finally, to log out, a user simply needs to call \texttt{logout} on the appropriate
\texttt{session} object: \texttt{logout(session)}.
\end{enumerate}

We chose to keep the \texttt{session} object external from the \texttt{Graffiti} class
to prevent developer mode errors, where an application attempts to perform an action
that requires authentication without asking the user to log in first.
These mode errors are detected by static type checking.
Separating the \texttt{session} object also allows an application to be logged in to
multiple actors at once, which can be useful for managing pseudonyms or anonymous interactions.

\subsection{Objects and CRUD}

Objects, as overviewed in Section~\ref{concepts:objects}, include
the following meta properties:
\begin{enumerate}
\item
\texttt{value} (object): The actual data, which can be any JSON object~\cite{json}.
\item
\texttt{url} (string): A string to uniquely identify and locate the object.
\item
\texttt{actor} (string): The URI of the actor that created the object.
\item
\texttt{channels} (array of strings): A list of channels that the object is contained in.
\item
\texttt{allowed} (array of strings or undefined): An optional list of actor URIs that are allowed to access the object.
When undefined, any actor can access the object if they know either its \texttt{url} or one of its \texttt{channels}.
\item
\texttt{revision} (number): A Unix time\-stamp, Lamport time\-stamp~{\cite{lamport}}, or other counter that
increases with each modification of the object.
\end{enumerate}

Subtly, \texttt{channels} and \texttt{allowed} do not necessarily
reflect \emph{all} the channels an object is contained in, or
\emph{all} the actors that are allowed access the object.
Like a BCC email, it can be useful to send a message to multiple parties
without revealing their existence to one another, preventing
context collapse.
So \texttt{channels} and \texttt{allowed} are both \emph{masked} to any
actor other than the object's owner so that they only show information an actor implicitly knows,
like the channel their application queried to discover the object.
However an application can always
add metadata, including channels or allowed lists, to the object's \texttt{value}
to reveal additional information.

Objects can be interacted with through familiar CRUD methods:
\texttt{put}, \texttt{patch}, and \texttt{delete} for modifying objects,
and \texttt{get} for retrieving objects.
The modification methods each require a \texttt{session} whereas
for \texttt{get} the \texttt{session} is optional,
enabling a user to view content without being logged in.
The API responds identically for an object that does not exist,
one that has been deleted, or one that the actor is not allowed to see,
so that no information about an object's status is leaked.

The \texttt{url} and \texttt{actor} are assigned to an object when it is first \texttt{put}
and are immutable.
As mentioned in our discussion of total reification in Section~\ref{concepts:total-reification},
only the actor that created an object can mutate the object at that \texttt{url}.
An object's owner can arbitrarily mutate the object's
\texttt{value}, \texttt{channels}, and \texttt{allowed} list.
The \texttt{revision} field is increased by the underlying implementation on each modification.

We expect that the ecosystem will contain many posts that a particular application cannot understand and will wish to ignore.  To simplify this process,
\texttt{get} requires the application developer to specify
a JSON schema, a widely used and flexible standard for describing
the structure of JSON objects~\cite{jsonschema}.
Any object that does not match the provided schema is rejected.
The API includes type inference on these schemas which makes it possible to work
with objects as if they were strongly typed, even though they can take on arbitrary values.

\begin{minted}{javascript}
const object = await graffiti.get(url, {
  value: {
    required: ["content"],
    properties: {
      content: { type: "string" },
    }
  }
}); // Error if the object does not match
const content = object.value.content; // string
\end{minted}

By default, JSON schemas also match objects that contain additional,
unspecified properties, allowing developers to freely
add new object properties without breaking
interoperability with existing applications,
encouraging the folksonomy.

\subsection{Discover}

\texttt{discover} lets an application developer query for objects
from a particular set of channels.
Objects are returned from \texttt{discover} as an \emph{asynchronous generator}.
This allows the application to process objects as they come in,
using a \texttt{for\ldots{await}} loop, rather than waiting for all objects
to be fetched. The generator also hides the potential complexity of
query pagination below the API.

Like \texttt{get}, \texttt{discover} can be given an optional \texttt{session}
to access private objects, and it requires a JSON schema.
The schema,
in addition to providing type safety, makes \texttt{discover} far more efficient
as a query mechanism, limiting the objects returned to only those that interest the particular application.
For example, the channel named by an actor's URI has many overloaded uses,
including being the site for the actor to broadcast Twitter-like posts,
and the site where they receive direct messages.
To receive only direct messages from Alice,
sent over the past day, a developer may do the following:

\begin{minted}{javascript}
const iterator = graffiti.discover(
  [session.actor], // Look in "my" channel
  {
    value: {
      required: ["content", "published", "to"],
      properties: {
        content: { type: "string" },
        published: {
          type: "number",
          minimum: Date.now() - 24 * 60 * 60 * 1000
        }
        to: { const: session.actor }
      }
    },
    actor: { const: "https://alice.example.com" },
  },
  session
)
for await (const object of iterator) {
  console.log("alice → me:", object.value.content);
}
\end{minted}

Reified activities can also be processed to construct more
complex \texttt{discover} schemas.
For example, the schema may exclude objects by
actors who have been ``blocked'' by a sufficient
number of other actors.
%DK really?  this seems like for more than a schema language, turning into a general graph-database query language?

The \texttt{discover} iterator will end once all objects
currently matching the channels and schema have been fetched.
It ends by returning a
\texttt{continue()} function,
which can be used to resume
the \texttt{discover} call.\footnote{
    A \texttt{cursor} string is also returned, which can be
    serialized for when an application closes and reopens.
    The \texttt{cursor} can then be passed into a \texttt{continueObjectStream}
    method and will produce the same result as \texttt{continue()},
    however it is does not preserve type safety and is less convenient.
}

\texttt{continue()} returns new objects, changed objects, \emph{and}
tombstones for objects that have been deleted or modified so that they no longer match the
specified channels, schema, or actor.
In total, these allow an application to keep an up to date state
without having to re-fetch all objects.
A tombstone only includes an object's URL and the time of deletion.
Tombstones must inevitably expire to preserve a user's right to be forgotten,
causing \texttt{continue()} to throw an error
that signals to the application to poll \texttt{discover} from scratch.

\texttt{continue()} can be recursively called for applications,
like direct messaging, that require real-time updates.
Below the API, an implementation might perform complex rate limiting on these recursive calls or
adaptively switch to a more efficient push-based transport, like a WebSocket, but this is hidden from developers working above.

\subsection{Recovery Methods}

It is intentionally impossible to get a global view of all objects
in Graffiti, to prevent context collapse. Most of the time,
an application must know either an object's URL or one of the channels
the object is contained in to retrieve it. However, we relax this
constraint for an actor's \emph{own} objects, to prevent them from
``losing'' objects.

The \texttt{recoverOrphans} method finds objects that an actor has created
that are not contained in \emph{any} channel, and \texttt{channelStats}
returns a list of all the channels that an actor has posted in.
Both methods output streams similar to \texttt{discover}
and require a \texttt{session}.
%DK ATP, how about instead allowing a null argument to discover which means that you are not looking in any channel but are instead looking through all your own posts (possibly filtering by a schema specification, which could for example indiciate that you only want posts not in any channel)

These methods are not intended to be used in most applications but are
necessary for creating tools like Facebook's ``activity log,''
or other novel digital hygiene tools,
that allow users to audit their system-wide usage.

\section{Above and Below}
\label{above-and-below}

We now describe the tooling we have built \emph{above}
the Graffiti API that makes development even easier, and
the implementations we have built \emph{below} the API
that realize different trade-offs between efficiency and
robustness.

While we have made insights on both fronts that at
the least demonstrate \emph{feasibility},
there are, of course,
many unanswered questions and unexplored ideas.
Fortunately, by having a fixed API in the middle, work done above the API
is \emph{decoupled} from work done below.
It is possible to drop in a new, more efficient, or privacy-preserving implementation
without upheaving all the tools and applications that have been built on top.

\subsection{Above}

While the API has hidden significant complexity below,
like the need to write and deploy server code or the need to
consider the server infrastructure at all,
writing front-end code is still inaccessible
to many people.
We have made some initial steps toward reducing the
programming complexity by allowing for the \emph{reactive} and
\emph{declarative} programming of Graffiti applications
and by providing a growing library of reusable design patterns.
We also point to a future where Graffiti applications could be built
without any code at all.

\subsubsection{Vue Plugin}

Vue is a popular front-end framework that enables ``reactive'' programming where
changes to the underlying state are automatically reflected in the UI.
We built a plugin for Vue\footnote{
  \url{https://vue.graffiti.garden/variables/GraffitiPlugin.html}
}
that provides a variant of
\texttt{discover} which returns a reactive array of objects rather than
an asynchronous stream.
Any CRUD operation that the user triggers locally instantly updates
the relevant array, while remote changes can be polled manually
with \texttt{poll()} or automatically
with an \texttt{autopoll} setting.

This reactivity is available in a traditional programming
environment as well as in a declarative HTML-like template.
The plugin maintains complete TypeScript support and
is a relatively light wrapper over a Graffiti-specific reactivity library we have developed\footnote{
   \url{https://sync.graffiti.garden/classes/GraffitiSynchronize.html}
}.
Therefore, support for other front-end frameworks, like React, will be
fairly easy to add in the future.

Below is the complete template for a micro social application,
built with the plugin, that shows a (continuously updating) list of all messages posted to a real-time group chat and enables the user to post their own message:
\begin{minted}{vue}
<form v-if="$graffitiSession.value"
  @submit.prevent="$graffiti.put({
    value: { content },
    channels: [ 'my-group' ]
  }, $graffitiSession.value)"
>
  <input v-model="content" />
  <input type="submit" value="Post" />
</form>
<button v-else @click="$graffiti.login()">
  Log In to Post!
</button>
<GraffitiDiscover autopoll
  v-slot="{ objects: messages }"
  :channels="[ 'my-group' ]"
  :schema="{
    properties: {
      value: {
        required: ['content'],
        properties: {
          content: { type: 'string' },
        }
      }
    }
  }"
>
  <ul>
    <li v-for="message in messages">
      {{ message.value.content }}
    </li>
  </ul>
</GraffitiDiscover>
\end{minted}

\subsubsection{Pattern Library}

We have started compiling a library of common patterns\footnote{
    \url{https://vue.graffiti.garden/examples/}
} that currently includes likes, profiles, following, threaded comments,
and different messaging patterns.
These patterns build on top of the Vue plugin
and are self-contained in \emph{single HTML files}
which can be downloaded, modified, and copied into other
applications.
Perhaps this format can encourage the copy and paste ``pimp my page''
programming that was once possible on MySpace~\cite{copypasteliteracy},
but with the ability to add new \emph{functionality} rather than just styling.

\subsubsection{Future Work}
\label{above-and-below:above:future-work}

There is certainly potential for more low or even no-code tools
to be developed on top of the Graffiti API. It is an open
question as to whether it is possible to build an intuitive and completely
no-code system for Graffiti, like a graphical editor, without limiting
its extensibility.

Additionally, in our own application development,
we found that inline AI tools like GitHub Copilot are already
capable of generating large portions of functional Graffiti code.
This is especially true when writing applications with the
Vue plugin, perhaps because of the reduction in boilerplate
code and the declarative nature of the templates.
It is not hard to imagine a future where a developer asks
a large language model to ``make a messaging app''
or to ``add a bookmark button to posts in this application'' and
it simply works.

Work could also be done to develop ``meta'' applications.
Users of these applications could easily change their functionality
by swapping out different plugins,
which could represent different feed sorting algorithms
(similar to \cite{threeleggedstool, bluesky}),
different moderation policies, different styles,
or different features.

Alternatively, Graffiti could serve as a new basis for
\emph{computational media}~{\cite{computationalmedia}},
where the boundary between developing and using applications is blurred.
This would require representing Graffiti applications with
Graffiti objects, potentially by extending the
collaborative editing capabilities we demonstrate in Section~{\ref{case-studies:wikiffiti}}.
Graffiti's social focus may fill gaps left by existing computational media
systems like WebStrates~{\cite{webstrates}} and its successors, CodeStrates~{\cite{codestrates}}
and Mirrorverse~{\cite{mirrorverse}}, which provide malleable interactive environments
for small, trusted teams, but have limited affordances
for \emph{moderation} or \emph{discovery} of social content.

\subsection{Below}

The Graffiti API is relatively simple
and so its implementation must solve just two main problems:
storing objects and discovering objects by channel.
One solution is to store all objects on a massive database
and then discovery is trivially a query on that database.
Unfortunately, this centralizes all control of the network
on that single server.

Instead, we explore \emph{decentralized} implementations of the API.
In the ``remote'' implementation, objects are served by a ``federation'' of databases,
each of which is a small version of the would-be centralized database.
In the ``commodity storage'' implementation, the problems of storage and discovery are split up,
with existing providers, similar to Dropbox, handling storage,
and a new entity, called a ``tracker,'' handling discovery.

These implementations have trade-offs, with
the former implementation being more efficient,
and the latter more robust.
Fortunately, multiple implementations can be understood within one \emph{meta}-implementation,
per Requirement~\ref{requirements:parallel-implementations},
allowing new implementations,
with better trade-offs, to be organically adopted without
invalidating all the data served by older implementations.

These parallel implementations are possible because, as we have mentioned, an object's URL starts with a scheme that indicates
how the various CRUD methods should resolve and interact with the object.
For example, if an actor tries to \texttt{delete} an object whose URL starts with the scheme
\texttt{graffiti:remote:}, their meta-implementation will employ the remote implementation,
while if the URL starts with the scheme \texttt{graffiti:cs:},
their meta-implementation will employ the commodity storage implementation.
To handle \texttt{discover}, the meta-implementation simply aggregates
the results of each implementation-specific \texttt{discover}
operation into one stream.

There are two additional questions we need to resolve to
make coexisting implementations work.
Which implementation should be used
when an actor \emph{first} \texttt{put}s an object?
And what \emph{identity} system will work across multiple
implementations?

\begin{figure}[h]
    \includegraphics[width=\columnwidth]{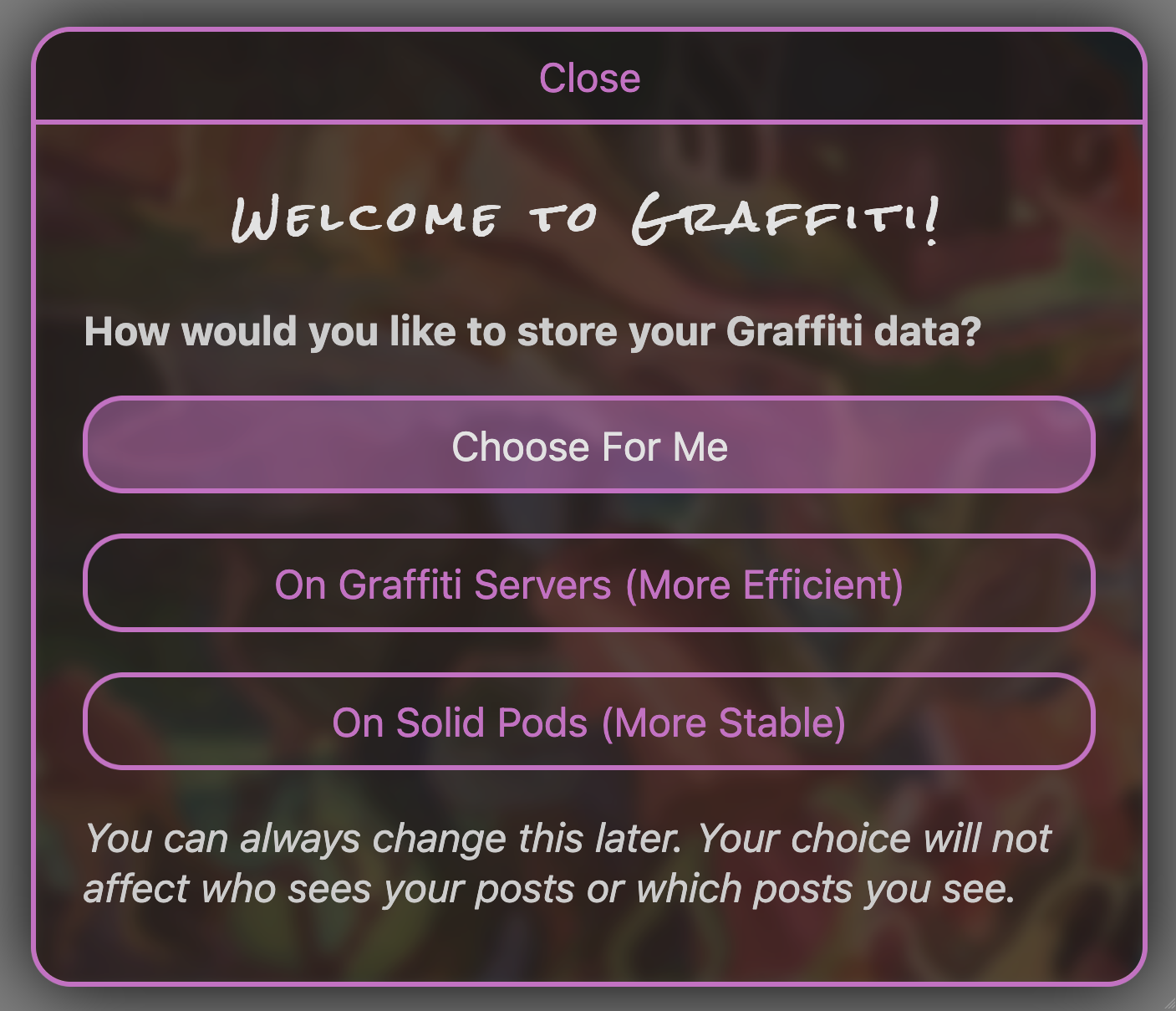}
    \caption{The modal that pops up when the user triggers a \texttt{put()} for the first time.}
    \Description{The figure shows a window titled "Welcome to Graffiti!" in a Graffiti-style font. Below a question asks "How would you like to store your Graffiti data?" and three buttons are presented "Choose for Me", "On Graffiti Servers (More Efficient)", and "On Solid Pods (More Stable)". The "Choose for Me" option is emphasized. Below there is a note: "You can always change this later. Your choice will not affect who sees your posts or which posts you see". There is also a close button at the top of the window.}
    \label{above-and-below:figure:choose-protocol}
\end{figure}

For the question of first \texttt{put}s, we simply give the user the choice.
The first time a user attempts to \texttt{put} an object,
a popup opens,
as shown in Figure~\ref{above-and-below:figure:choose-protocol},
asking them which implementation they would like to use,
as well as implementation-specific information, such as what server
they want to store their data on.
We have tried to make the experience relatively frictionless, with
reasonable defaults for
users who do not know or care about the underlying infrastructure.
Importantly, this interface is built into the \emph{meta-implementation}
of the API and so it is not the concern of developers building
applications on top of the API.

As for identity, we currently use a decentralized protocol called Solid OIDC~\cite{solidoidc}.
It has the useful property that a user only needs to log in once
with an ``identity provider'' which then silently authenticates
them with other services.
It is also already used by the Solid protocol~\cite{solid},
which we use in our commodity storage implementation as a ``decentralized Dropbox.''

Unfortunately, Solid OIDC is not a widely deployed and familiar identity solution,
like ``Log in with Google.''
We are keeping an eye on the Decentralized Identifier (DID) specification, which attempts to unify
various identity methods, but, currently, DIDs
do not capture Solid OIDC's property of silent authentication~\cite{dids}.
This would produce a tedious and confusing experience analogous to having to log in every time you receive email from a new server.
We hope that DIDs or new specifications evolve past this restriction.

After discussing our two main implementations, we discuss
a bonus ``utility'' implementation that demonstrates
the additional power of an abstract API.

\subsubsection{Remote Implementation}
\label{above-and-below:remote-protocol}

The remote implementation involves a federation of database servers,
each of which exposes the Graffiti API through an HTTP API.
Like email, users can choose which server they would
like to use or run their own.
Unlike email, the servers do not send objects to each other.
Instead, the client fetches objects directly from servers,
selecting which server to fetch from based on a \emph{domain}
in the object's URL, just as different implementations are selected by \emph{scheme}.

For discovery, we currently provide a registry of servers
and clients call \texttt{discover} on all servers in the registry.
If many servers are deployed, discovering from every one
will be infeasible, and so instead we can use an ``announce''
protocol, detailed in Appendix~\ref{appendix:announce}. The protocol builds on top of the remote implementation,
but only requires the client to
call discover on the servers containing relevant data,
plus a small constant number of well-known
``\emph{rendez-vous}'' servers.

Our server-side software for the remote implementation
consists of a NestJS server with a CouchDB database,
bundled as a Docker image for easy cross-platform
deployment\footnote{
\url{https://github.com/graffiti-garden/implementation-remote}
}.

\subsubsection{Commodity Storage Implementation}
\label{above-and-below:commodity-storage-protocol}

An alternative Graffiti implementation builds on top of existing
\emph{commodity storage} providers,
similar to Dropbox.
Like the remote implementation, users get to choose which storage
service they would like to store their objects in, including
from one they host themselves.
The client software
manages channel-specific files on the storage provider,
each containing all the objects the actor has published to that channel.
The client software shares links to those channel files with a \emph{tracker}.
The tracker is similar to a BitTorrent tracker~\cite{bittorrent} but
rather than mapping torrent IDs to IP addresses
it maps channel strings to channel file URLs.
To run \texttt{discover}, the client software queries the tracker
about each channel of interest and then fetches each of the resulting URLs for objects.

The commodity storage implementation is less efficient than the remote implementation because each
\texttt{discover} call requires a network request per-channel-per-actor,
where as the remote implementation requires only one request per remote server.
If, like email or Mastodon, many actors, especially those in regular contact,
use the same server~\cite{mastodonchallenges},
then \texttt{discover} on the remote implementation can take as
few as \emph{one} request, while requests made by
the commodity storage implementation scale with the number of actors.
Additionally, the schema provided to \texttt{discover} cannot be used for any
server-side filtering---a client must download \emph{all} objects in a channel
and discard ones that do not match the schema.

Still, for users who tend to participate in small networks, these issues
of scale are not a concern and the use of ``simple'' storage servers can be appealing.
The user does not need to worry that an experimental new service
will go offline or lose their data, and we, as implementers do not need to
worry about developing the infrastructure.

At the time of writing, we have built the tracker server and client\footnote{
  \url{https://github.com/graffiti-garden/link-service}\\
  \url{https://github.com/graffiti-garden/link-service-client}\\
  WebTorrent was considered instead of a custom tracker,
  but unfortunately it does not persist announcements more than a few hours.
}, and the complete implementation software,
using Solid ``pods'' as storage providers,
will soon be complete.

\subsubsection{Local Implementation}

\begin{figure*}[t]
    \includegraphics[width=\textwidth]{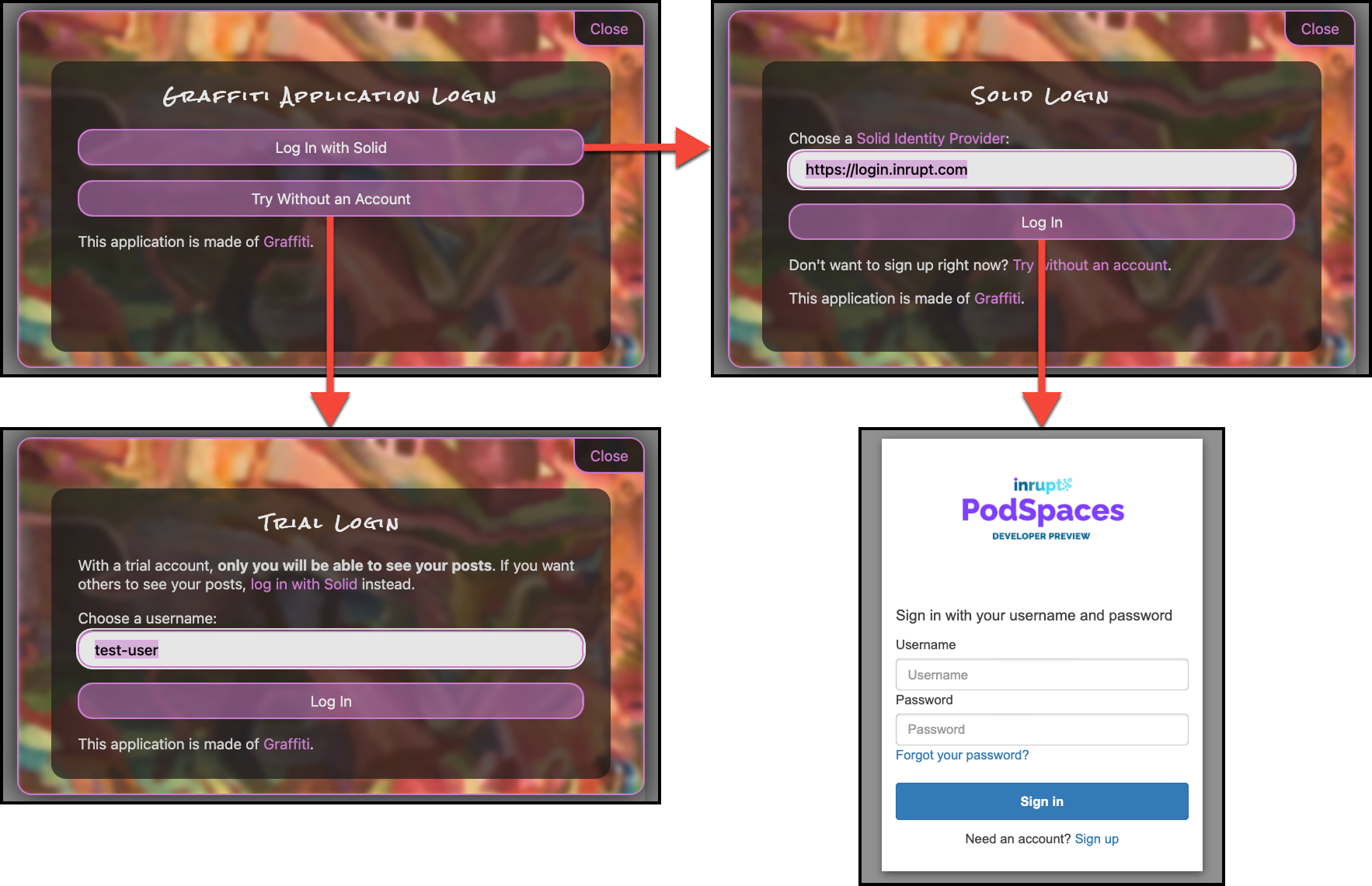}
    \caption{The flow that pops up when the user triggers a \texttt{login()}.}
    \Description{The figure shows a series of pop-up windows with arrows between them. The first window says "Graffiti Application Login" and below there are buttons to "Log In with Solid" and "Try Without an Account". An arrow pointing from the "Try Without and Account" option points to another window titled "Trial Login" which contains the paragraph "With a trial account, *only you will be able to see your posts*. If you want others to see your posts, log in with Solid instead" and below that a form saying "Choose a username" with "test-user" entered in the input and finally a button to "Log In". From the "Log In with Solid" button in the original panel there is an arrow pointing to a window titled "Solid Login" which says "Choose a Solid Identity Provider" and below an input containing "htttps://login.inrupt.com" and "Log In" and finally "Don't want to sign up right now? Try without an account." An arrow points from the "Log In button of the "Solid Login" window to a new window that looks minimal rather than the more maximalist graffiti designs of the previous windows. This minimal window is titled "inrupt PodSpaces Developer Preview" with the paragraph "Sign in with your username and password" and below a username and password form, text that says "Forgot your password", a "Sign in" button, and finally "Need an account? Sign up".}
    \label{above-and-below:figure:login}
\end{figure*}

This implementation runs entirely client-side
and handles objects with the URL scheme \texttt{graffiti:local:}.
Any data created using the local implementation is only visible on the device on which
it was created.
While such a restriction may seem like a non-starter for an implementation
of our \emph{social} API, it is invaluable for development.
It provides a similar functionality to running a \texttt{localhost} development server,
but
does not require the complexity of standing up a local server.
Developers can easily create test actors and test objects on the local implementation
without polluting their own online presence or existing Graffiti applications.
Finally, the implementation is accessible in deployed applications
and lets users try a Graffiti application out without
having to create or log in to an account, as shown in
Figure~\ref{above-and-below:figure:login}.

Because of its relative simplicity, the local implementation\footnote{
    \url{https://github.com/graffiti-garden/implementation-local}
} serves as the reference implementation of the Graffiti API.

\subsubsection{Future Work}
\label{above-and-below:below:future-work}

The remote implementation offloads object filtering work to complex storage servers,
while the commodity storage implementation shifts that work to clients, keeping the servers simple.
A middle ground is possible by adding more structure to the organization of objects on
the storage servers, and by adding complexity to the tracker to index that structure.
For example, rather than putting all objects in a channel into one file,
those files could be subdivided according to common object schemas.
This would maintain the use of simple servers, but make \texttt{discover} calls more efficient.

In the commodity storage implementation, the tracker also serves as a central point of control and so, like in BitTorrent,
it could be replaced with a distributed hash table~\cite{bittorrentdht}. In small networks,
a tracker may not be necessary at all and can be replaced with a gossip protocol.

Currently each server owner can see all the objects and channels
that pass through it. While federation allows users to switch
to servers they trust, those servers are still vulnerable to
subpoenas, putting groups, like activists, at risk.
It would be valuable to build end-to-end encrypted versions of the remote servers,
tracker, and commodity storage servers, perhaps building on
searchable encryption~\cite{searchableencryption}.

One vulnerability of any system with portable login,
including systems like Solid and the AT Protocol~{\cite{bluesky}},
is the potential of logging into
a malicious application that could steal your data or
post on your behalf without consent.
This risk may make users hesitant to try new applications,
restricting the ecosystem.
Future work should introduce login scopes, such as
the ability to grant an application the permission only to read objects
that match a particular schema and reside in a particular set of channels.
Since the commodity storage implementation bootstraps off of
existing storage services that lack such scoping,
it may be necessary to implement scoping client-side
via an iframe to a trusted intermediary domain.

\section{Case Studies}
\label{case-studies}

Our case studies demonstrate the diversity of applications
that can be built on top of the Graffiti API and explore interoperation
between those applications.
The studies also demonstrate some of the unusual design patterns that
Graffiti makes possible when the boundaries between applications
are fuzzy and no one is really in control.

Other than the interoperation we describe, no \emph{unexpected}
interoperation occurs, thanks to our channels concept.
For example, group chat messages do not show up as posts on
a microblogging site even though both use similar object schemas,
because they are each posted to different channels.

All of the applications are written with pure client-side code on
top of the Graffiti Vue plugin.

\subsection{Glitter and The Glue Factory}
%DK Why lump these together?   Describe Glitter first (and discuss how easy it is to implement) than describe Gloof and explain how they interoperate

The following example demonstrates how a community-specific application,
%DK I would call this a social site rather than an application (and explain why).
\emph{The Glue Factory}\footnote{
\url{https://gluefactory.live}\\Source: \url{https://github.com/horseyhouse/glue-factory}
}, can interoperate with a general-purpose application, \emph{Glitter}\footnote{
\url{https://glitter.graffiti.garden}\\Source: \url{https://github.com/graffiti-garden/glitter}
},
without the explicit permission of Glitter.
Not all of the community-specific features on The Glue Factory
translate to Glitter, and not all of the interconnectedness on Glitter
is forced upon The Glue Factory.
Still, there is enough interoperation between the two to be meaningful.

\emph{Glitter} is a text-centric microblogging platform
where users can post, follow, reply, and change their display name.
Replies are threaded and, like Twitter, appear in the replier's followers' feeds, in addition to the reply thread.
There is also a directory that users can add themselves to,
so others can find them.
%DK Is the directory glitter specific?   why?   finding users seems important for most applicatoins.

\emph{The Glue Factory} is an application made for a local
venue to share event flyers.
The flyers are displayed in a grid, similar to Instagram,
but there is just one feed, collectively curated by
the application's fixed set of venue organizers.
Flyers can be replied to, but the organizers may remove replies they disapprove of.

Importantly, The Glue Factory lets
top-level repliers ``crosspost'' their replies
to Glitter, effectively sharing the flyer like a Quote Tweet.
Glitter does not support images and so only the
event description appears.
Replies to the crosspost made on Glitter also appear on The Glue Factory
and vice versa, \emph{unless} The Glue Factory's organizers ``remove'' the reply,
in which case it only appears on Glitter, regardless of where it originated.
Additionally, replies to the crosspost made on Glitter appear in the replier's followers'
feeds while replies made on The Glue Factory only appear in the reply thread,
similar to the difference between Instagram-like and X-like replies described in Section~\ref{concepts:channel-replies}.
Some of this interoperation is shown in Figure~\ref{case-studies:fig:gloof-and-glitter}.

\begin{figure*}[t]
    \centering
    \includegraphics[width=\textwidth]{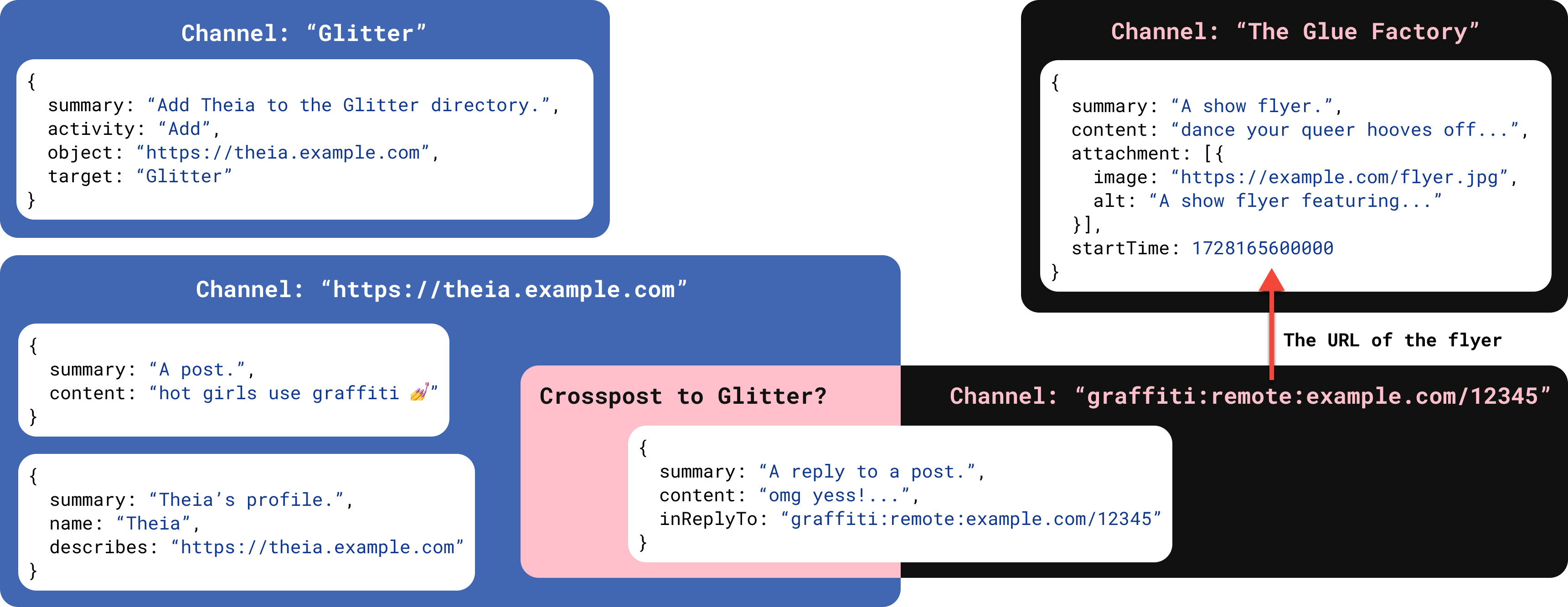}
    \Description{
     The figure shows a series of four rounded rectangles, some overlapping like a Venn diagram, each containing one or multiple JSON blobs. Each rectangle contains at the top a header that says "Channel" followed by a string in quotes and each JSON blob contains a "summary" property. The first rectangle has the channel "Glitter" and contains one JSON blob with the summary "Add Theia to the Glitter directory". The rest of the blob contains { activity: "Add", object: "https://theia.example.com", target: "Glitter" }. The second rectangle has the channel "The Glue Factory" and contains one JSON blob with the summary "A show flyer." The blob also contains { content: "dance your queer hooves off...", attachment: [{ image: "https://example.com/flyer.jpg", alt: "A show flyer featuring..." }], startTime: 1728165600000 }. A third rectangle has the channel "https://theia.example.com" and contains two full JSON blobs and part of one blob that is overlapping with the last rectangle. The first blob has the summary "A post." and one addition property "content" with the value "hot girls use graffiti [nail polish emoji]". The second blob has the summary "Theia's profile." and also contains { name: "Theia", describes: "https://theia.example.com" }. The final blob contains the summary "A reply to a post." and also { content: "omg yess!...", inReplyTo: "graffiti:remote:example.com/12345" }. That blob is contained in a rectangle with the channel "graffiti:remote:example.com/12345". There is an arrow pointing from the channel name of the last rectangle to the blob in show flyer blob in the second rectangle. The arrow is labeled "The URL of the flyer". The overlap between the rectangles with the channels "https://theia.example.com" and "graffiti:remote:example.com/12345" is highlighted and labeled "Crosspost to Glitter?"
    }
    \caption{Example object values and channels used by Glitter and The Glue Factory.
    When crossposting is selected, the reply is put into a Glitter-known channel in addition to a Glue Factory-known one.}
    \label{case-studies:fig:schemas-and-channels}
\end{figure*}

The applications make use of the objects and channels
shown in Figure~\ref{case-studies:fig:schemas-and-channels}.
Importantly, both applications use the same reply object schema,
and the post object schema used by Glitter is a subset of the flyer
object schema used by The Glue Factory.
This overlap makes interoperation \emph{possible},
but selecting ``crosspost to Glitter'' in The Glue Factory intentionally collapses the
channel usage of the two applications,
causing the interoperation to actually \emph{happen}.

One additional note is that flyers posted to The Glue Factory include a \texttt{startTime}
to display event dates.
This metadata is also used to populate a separate calendar application
used by the co-operative that owns the venue space.

\subsection{Parallax and Provenance}
\label{case-studies:parallax}

%DK perhaps before getting into the crazy bits, talk about how easy it is to implement a basic group chat application ("Grack").   Each group is defined by a channel, and you invite someone to the group by notifyig them about the channel.  There is no admin---anyone in the group can invite others---and no way to remove people without their consent (but you can always create a new group excluding them).  Maybe also describe how you would design a facebook clone ("Gracebook") differently: a user can post content for friends by putting it in their own actorID channel but with an allowlist consisting of their friends.  alternatively they can create a different channel for sharing posts to their friends and invite all friends to subscribe to it.

\begin{figure*}[t]
    \centering
    \includegraphics[width=\textwidth]{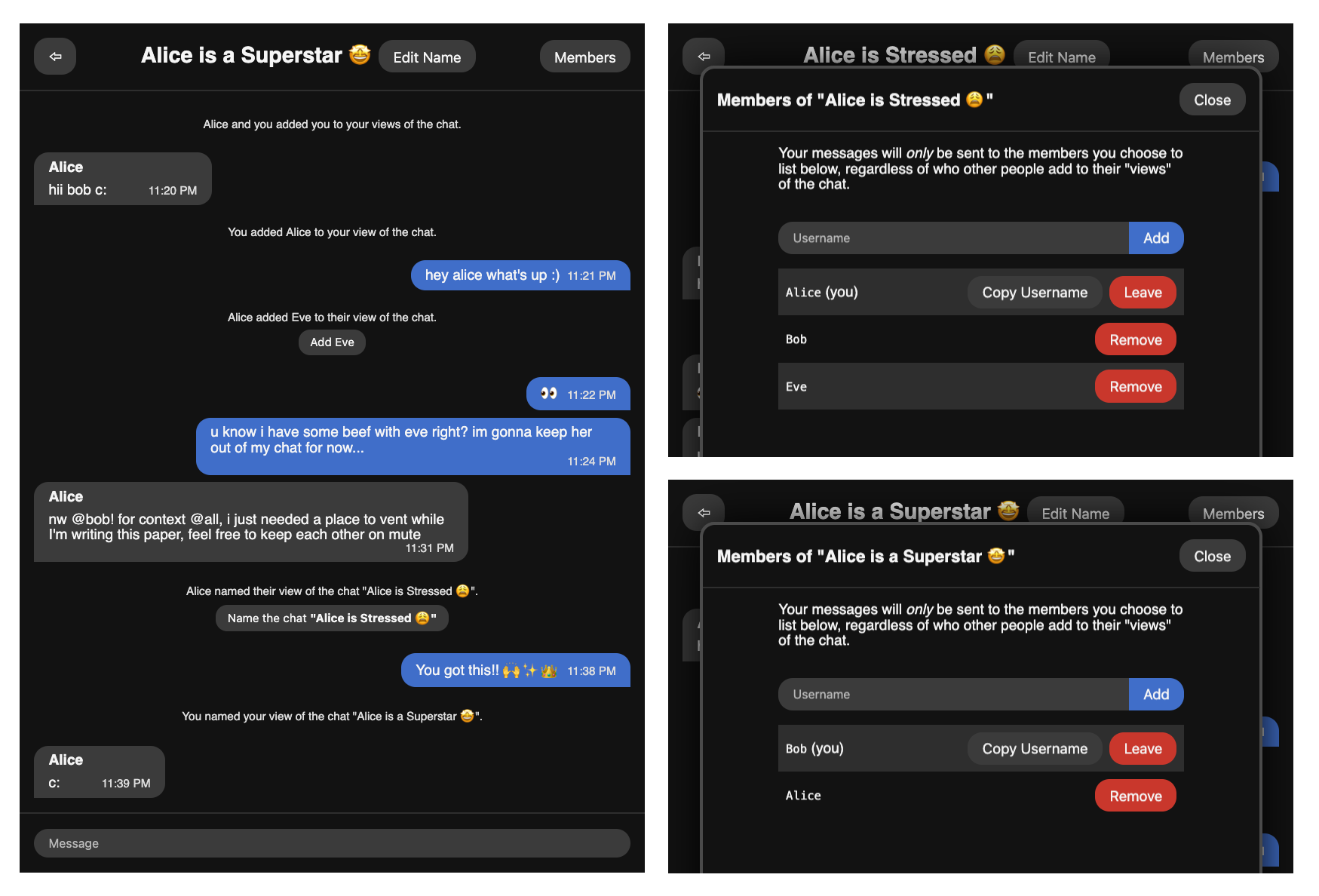}
    \caption{An interaction on Parallax. On the left is Bob's view of an exchange. On the right are Alice and Bob's \emph{different} membership lists.}
    \Description{
     The figure shows a screenshot of a chat application and cropped screenshots of the "Members" popup in that application. The chat window is titled "Alice is a Superstar [starstruck emoji]" with a button "Edit Name" and a button "Members". Below that the transcript contains both centered text and text in bubbles in either side. Center text: "Alice and you added you to your views of the chat". Left bubble: "Alice: hii bob c:". Center text: "You added Alice to your view of the chat". Right bubble: "hey alice what's up :)". Center text: "Alice added Eve to their view of the chat." and below a button labeled "Add Eve". Right bubble: "[side eye emoji]". Right bubble: "u know i have some beef with eve right? im gonna keep her out of my chat for now...". Left bubble: "Alice: nw @bob! for context @all, i just needed a place to vent while I'm writing this paper, feel free to keep each other on mute". Center text: "Alice named their view of the chat 'Alice is Stressed [weary face emoji]'" and a button "Name the chat 'Alice is Stressed [weary face emoji]'". Right bubble: "You got this!! [raising hands emoji][sparkle emoji][crown emoji]". Center text: "You named your view of the chat 'Alice is a Superstar [starstruck emoji]'". Left bubble: "Alice: c:". Finally there is a message text box. On the right one popup reads "Members of 'Alice is Stressed [weary face emoji]'" with the text "Your messages will *only* be sent to members you choose to list below, regardless of who other people add to their 'views' of the chat." below there is place to enter a username a button "Add" and then finally a list containing "Alice (you)", "Bob", "Eve" with the labels "Leave", "Remove", "Remove" respectively. On the right one popup reads "Members of 'Alice is a Superstar [starstruck emoji]'", containing the same text and form but the members "Bob (you)" and "Alice" with the buttons "Leave" and "Remove" respectively.
    }
    \label{case-studies:fig:parallax}
\end{figure*}

\emph{Parallax}\footnote{
\url{https://parallax.graffiti.garden}\\Source: \url{https://github.com/graffiti-garden/parallax}
} is a real-time group chat application that demonstrates
how, under total reification, it is possible for \emph{every} user to employ a different
moderation scheme.
Specifically,
every user, from their own perspective,
is the sole administrator of \emph{all} group chats (that they know about) with
unilateral control over each group's name and membership.
The messages a user sends in a group can only be seen by the users they explicitly
put on that group's membership list.
However, users can also see the changes that other users
make to their own ``views'' of a group and are given the option
to \emph{voluntarily} incorporate those changes,
as shown in Figure~\ref{case-studies:fig:parallax}.

Under the hood, a group is represented by a random identifier,
generated when the group is created, and also used as the group's channel.
A change to a group's name is similar to a profile name change on Glitter, only it
\texttt{describes} the group identifier.
A group's membership can be changed with reified \texttt{"Add"}
and \texttt{"Remove"} activities.
Messages use the same object schema as posts on Glitter,
only they have \texttt{allowed} lists, which are determined
by aggregating the user's own \texttt{"Add"} and \texttt{"Remove"} activities
to determine the group membership state.

Of course, complete independence is not always desirable:
work usually done by just one group administrator must, in Parallax,
be done by every single group user.
We use Parallax to demonstrate an extreme,
but it can be
transformed into a more reasonable (but more restrictive) application called \emph{Provenance}\footnote{
\url{https://provenance.graffiti.garden}\\Source code is in the Parallax repository.
},
where a group's administrator is the \emph{creator} of the group chat.

Parallax and Provenance both interoperate,
but some messages sent from one will not be seen in the other
according to their unequal membership lists. However, this messiness is already present and tolerable
in messaging applications like Signal, where users can block other group members,
and email, where any reply can be sent to a different set of recipients.
There is exciting work to be done learning what interfaces make
Graffiti's inevitable asymmetry most accessible and engaging.

\subsection{Wikiffiti}
\label{case-studies:wikiffiti}

\begin{figure*}[t]
    \centering
    \includegraphics[width=\textwidth]{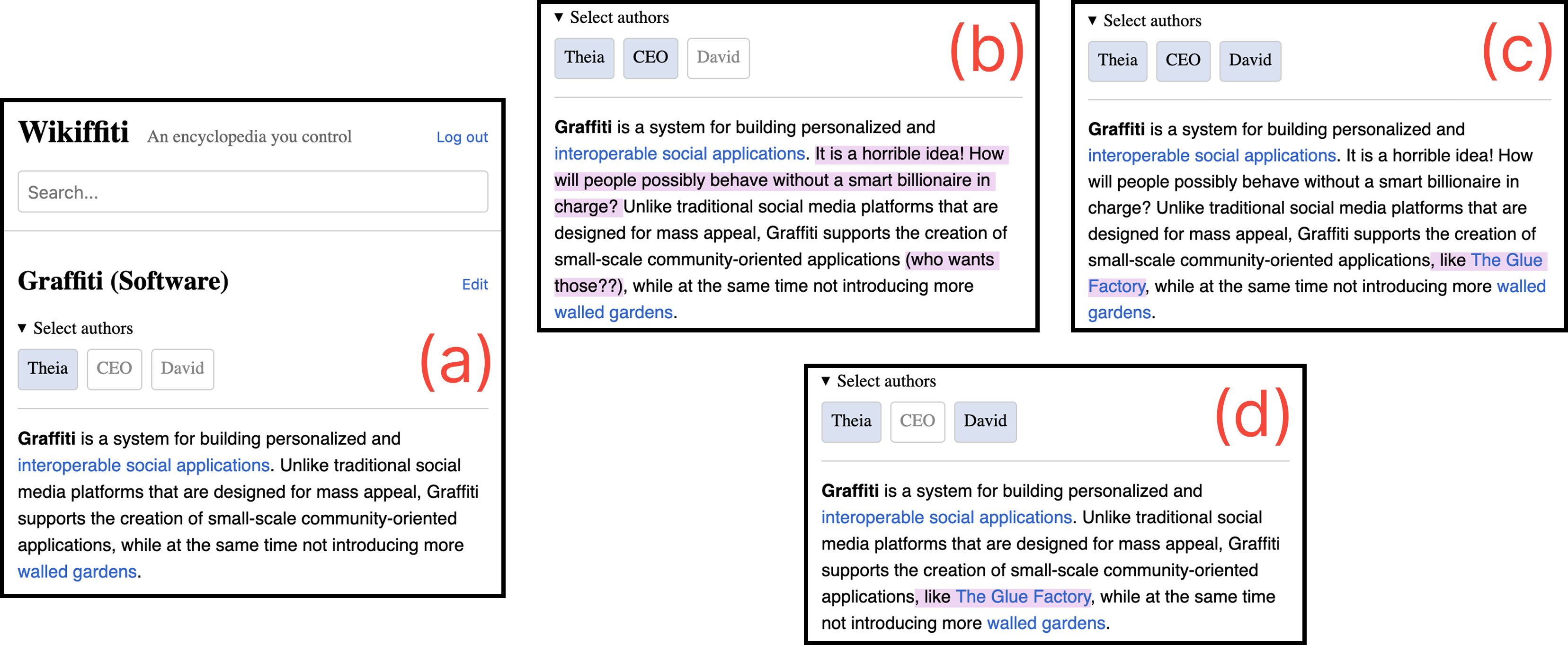}
    \caption{
    An interaction on Wikiffiti, with changes highlighted for clarity.
    (a) Theia creates a new Wikiffiti page. (b) CEO vandalizes the page. (c) David manually overwrites some of
    CEO's content. (d) All of CEO's edits are retroactively removed.}
    \Description{
    The figure displays four screenshots of a Wikipedia-like application labeled (a) through (d).
    (a) shows the header "Wikiffiti: An encyclopedia you control" with a "Log out" button and a "Search bar". Below, is the title "Graffiti (Software)" a button that says "edit" and an open dropdown listed "select authors" with three buttons labeled "Theia", "CEO" and "David". "Theia" is highlighted, while CEO and David are dimmed. Below, there is the text "Graffiti is a system for building personalized and interoperable social applications. Unlike traditional social media platforms that are designed for mass appeal, Graffiti supports the creation of small-scale community-oriented applications, while at the same time not introducing more walled gardens..
    (b) shows a similar partial interface to that of (a), continuing from the "select authors" on. Now "Theia" and "CEO" are highlighted but "David" is still dimmed. Some of the text is highlighted which we put in brackets. The text reads "Graffiti is a system for building personalized and interoperable social applications. [It is a horrible idea. How will people possibly behave without a smart billionaire to tell them what to do?] Unlike traditional social media platforms that are designed for mass appeal, Graffiti supports the creation of small-scale community-oriented applications [(who wants those??)], while at the same time not introducing more walled gardens.".
    (c) shows the same interface as (a) but now with all buttons selected, "Theia", "CEO", and "David". The text, with bracketed highlights reads "Graffiti is a system for building personalized and interoperable social applications. It is a horrible idea. How will people possibly behave without a smart billionaire to tell them what to do? Unlike traditional social media platforms that are designed for mass appeal, Graffiti supports the creation of small-scale community-oriented applications[, like The Glue Factory], while ensuring these communities do not turn into walled gardens."
    (d) shows the same interface as (a) but now with, "Theia" and "David" selected and "CEO" unselected. The text, with bracketed highlights reads "Graffiti is a system for building personalized and interoperable social applications. Unlike traditional social media platforms that are designed for mass appeal, Graffiti supports the creation of small-scale community-oriented applications[, like The Glue Factory], while at the same time not introducing more walled gardens."
    }
    \label{case-studies:fig:wikiffiti}
\end{figure*}

Wikiffiti\footnote{
\url{https://wikiffiti.graffiti.garden}\\Source: \url{https://github.com/graffiti-garden/wikiffiti}
} is a Wikipedia-like application that demonstrates that
collaborative editing in Graffiti is possible,
even though an actor can only mutate their own objects.
Additionally, unlike Wikipedia,
every user on Wikiffiti can choose which other users have ``permission'' to edit an article,
\emph{retroactively} undoing edits by unpermissioned users,
as shown in Figure~\ref{case-studies:fig:wikiffiti}.
The user can only control which users' edits \emph{they} see, not exert any global control---so different users may end up seeing different pages.

Edits to each Wikiffiti article are published to
the channel represented by the article's title.
This allows for basic exact-match lookup on titles and, like on Wikipedia,
``disambiguation'' pages can serve as manual search indexes when necessary.

Edits are published and composed together according to Logoot,
a conflict-free replicated data type (CRDT)~\cite{logoot,crdts}.
Logoot, and CRDTs in general, were developed for asynchronous collaborative editing
but, luckily for us, Logoot produces reasonable results when
some edits are ``dropped,'' as we do here intentionally.
Currently, our implementation is inefficient with a 40x space blowup;
however, there are plenty of existing optimizations that one could apply~\cite{logootbetter}
and release as part of a standard collaborative editing library.

Like Parallax, Wikiffiti is an extreme. Clearly not every user
has the expertise or desire to vet all the editors of
every article they read. In reality, the work of approving editors
or individual edits will be delegated to
a hierarchy of user access levels,
friend-of-a-friend networks of trust,
or automatic vandalism detectors, for example.
Still, the data underneath can always be reinterpreted, allowing
for new systems to independently evolve that
might be more welcoming to newcomers~\cite{wikibourgeoisie, wikirisedecline}
or promote edits made by women and non-binary people~\cite{wikigender}.
An application could even highlight edits that are vandalism to some,
but art to others: graffiti.

\section{Related Work}
\label{related-work}

Graffiti, so far as we know, is the only system that makes it
relatively easy to build such a wide variety of interoperating social applications.
Below, we discuss \emph{frameworks} that make it easy to build
\emph{non-\allowbreak{}interoperating} social applications
and \emph{protocols} that support interoperation but
make application development difficult and restrict the design space.
We also discuss other systems with related designs to Graffiti.

\subsection{Frameworks}

Facebook groups, subreddits, and
Slack workspaces are \emph{no-code} frameworks for creating social
applications.
Of course, these sub-applications have limitations,
like a strict user-moderator hierarchy
and predefined feature sets.
However, Reddit and Slack allow moderators to deploy
custom \emph{bots} that enable the
``delta'' ranking system in the subreddit \texttt{r/Change\allowbreak{}My\allowbreak{}View}
~\cite{changemyview}
and the PolicyKit moderation system in Slack~\cite{policykit}.

Déjà Vu is a more expressive framework that allows developers
to declaratively construct a social media application out
of a catalog of primitive \emph{concepts}.
Unfortunately, these concepts are limited in their extensibility.
For example, the ``scoring'' concept cannot be expanded
to support multiple ``reactions''
without creating a new concept and new server code~\cite{dejavu}.

Commercial frameworks, like Google Firebase, allow developers
to build on top of Google's database and identity systems,
enabling pure client-side development.
Firebase plugins for the front-end framework Vue and the low-code tool
Mavo~\cite{mavo} enable developers to write these applications declaratively.
For example, Mavo Chat\footnote{
    \url{https://dmitrysharabin.github.io/mavo-chat/}
} is a real-time chat application written in 156 lines of HTML
with Mavo and Firebase.
%DK could a graffiti remote server be implemented on firebase?

\subsection{Protocols}
\label{related-work:protocols}

Before 2018, social protocols, like Email and IRC,
and later diaspora*~\cite{diaspora} and Secure ScuttleButt~\cite{scuttlebutt},
were designed to support \emph{specific} application types:
messaging for the former and microblogging for the latter.
Since the introduction of ActivityPub~\cite{activitypub},
a new generation of social protocols have
been developed that, technically, support a wider variety of applications.
However, in practice,
building applications on these protocols can be limited by technical
difficulty, barriers to interoperability, or so much interoperability
as to cause context collapse.

All of the protocols we describe below are \emph{federated},
lacking a central point of control or,
at the very least, offering
a ``credible exit'' if one organization in the federation becomes
untrustworthy~\cite{howdecentralizedisbluesky}.

\subsubsection{ActivityPub}
\label{related-work:activitypub}

ActivityPub~\cite{activitypub} is the federated protocol underlying the
``Fediverse,'' which includes the applications
Mastodon (Twitter-like), Pixelfed (Instagram-like), and
Lemmy (Reddit-like).
Graffiti builds on ActivityPub's representation of social artifacts and
activities as extensible objects.
However, some actions in ActivityPub, like moderation, are not
reified into activities, or are
limited in their extensibility by server ``side effects''.

ActivityPub is organized into ``instances'' that conflate
a user's community, moderation service, application,
identity provider, and storage provider into one entity.
Moderation decisions by instance moderators are not reified activities
but actually delete or modify the target content.
This prevents their users from choosing a different moderation
service without leaving their instance and community
behind.
It \emph{also} means that federated content
needs to be moderated by \emph{every} receiving instance,
overwhelming already resource-constrained moderators, causing many
to resort to coarse moderation techniques like ``defederation''
\cite{securingfederatedplatforms, blocklistboundary}.
Note that this redundant labor is not necessary with totally
reified moderation as moderation actions are their own objects
that can be shared across applications.

ActivityPub distributes content using an ``actor model,''
where every actor has an inbox for receiving direct messages, like email,
and an outbox for broadcasting messages, like RSS.
While the actor model is conceptually elegant, it makes it difficult
to implement features as simple as replies.
A replier must send their reply to the original poster's instance which
triggers a server ``side effect'' that forwards the reply to the original
poster's outbox so their followers can discover it.
%DK this means the origial poster can block any replies they don't like from beig received by others
In practice, there is a flow diagram of at least 8 edge cases to consider
to prevent ``ghost replies''~\cite{stateofmastodon}.
Server side effects are needed to facilitate any interaction not centered around identity,
including all of those listed in Table~\ref{concepts:channels-and-applications}.
Additionally, side effects mitigate the autonomous extensibility
(see Requirement~\ref{requirements:autonomous-extensibility})
of ActivityPub
because all involved servers need to implement a side effect before
its corresponding interaction can be reliably used.

Thus there is no ``generic'' ActivityPub server
and custom-built servers
for Mastodon, Lemmy, and PixelFed each consist of hundreds
of thousands of lines of code.
Projects like Hometown, Smalltown, and Glitch\footnote{
    \url{https://glitch.com/fediverse}
} attempt to lower the barrier
to instance creation, but beyond basic configuration panels,
new application designs require modifying the underlying server code~\cite{smalltown,runyourownsocial}.

\subsubsection{Matrix}

The Matrix protocol\footnote{
    \url{https://spec.matrix.org/}
}
provides similar features to IRC and XMPP but content within a ``room''
can be stored across multiple servers and is end-to-end encrypted.
While generally intended for messaging, we mention Matrix because of a
prototype microblogging application built on top of it called \emph{Cerulean}~\cite{cerulean}.

Cerulean creates a ``timeline'' room for each of its users.
Users publish posts to their own timeline room
and follow each other by subscribing to each other's timeline rooms.
When a user posts, their application creates a
public ``thread'' room and replies are copied to both the
thread room and the replier's timeline room.
Matrix's client-server API is flexible enough that this prototype
was written with front-end code over a generic Matrix server.

This is similar to Graffiti's channel-based implementation of microblogging, described
in Section~\ref{case-studies:fig:gloof-and-glitter}.
One difference is that the ``thread'' and ``timeline'' rooms
in Cerulean
must be explicitly created by the poster.
This gives them automatic moderator privileges
under Matrix's rigid ``power level'' moderation
scheme, unlike channels
which are ``permissionless'' but allow arbitrary
moderation structures to be added on top via
reification.
This method also requires that someone is clearly responsible for instantiating the room,
which might not be the case for conversations centered around topics, external media, or locations.

Still, Cerulean's existence indicates that
Matrix might be modified to support the Graffiti API.

\subsubsection{The AT Protocol and Nostr}

The AT Protocol~\cite{bluesky} and Nostr\footnote{
\url{https://nostr.com}
} are two similar federated protocols,
with the AT Protocol being notable for underlying Bluesky.
Both protocols use a combination of ``personal data stores'' and ``relays''
to construct a queryable stream of \emph{all} data in the system.
Within the AT Protocol, this global stream is called the ``firehose''
and intermediate ``app views'' can consume content from this firehose to
construct algorithmic feeds, reducing client-side computation compared
to our own approach, at the cost of standing up custom servers.

A globally queryable stream of data makes contextual separation impossible.
For example, it is impossible for a user, Alice, to post a reply on either protocol
without that reply surfacing in a query for ``anything by Alice.''
This also appears to be causing inter-application collapse, similar to
what we predict in our discussion of Requirement~\ref{requirements:context-differentiation}.
An application called Flashes, built on the AT Protocol,
currently places all images a user has posted to Bluesky within
an Instagram-like grid. Users who want their Flashes feed to be
more curated are simply suggested to use a different account\footnote{
    \url{https://www.youtube.com/live/B7OwcXCE5Rg?t=1655s}
}.

The AT Protocol and Nostr both employ ``stackable
moderation,'' where users can opt-in to various ``labeling'' services.
Labeling-as-moderation is an approachable model that reflects existing expectations of a
top-down moderator-user hierarchy,
hence why we began our discussion of total reification
with \texttt{"Remove"} labels in Section~\ref{concepts:total-reification}.
However, labeling alone does not cover the more general democratic patterns
of moderation that we go on to describe are possible with total reification,
including the reified group membership and document authorship
that we demonstrate in Section~\ref{case-studies}.
Additionally, labelers on the AT Protocol need to run their own labeling servers.

Both protocols include extensible objects, but these objects are not
\emph{autonomously} extensible (see Requirement~\ref{requirements:autonomous-extensibility})
because the relays will not index objects with properties that are unfamiliar to them.
Additionally, both protocols sign all objects with their users' public keys, making
all content \emph{unrepudiable}, violating our Requirement~\ref{requirements:forgiving}.
Finally, the AT Protocol does not currently allow for
any private interactions, including private messaging.

\subsection{Adversarial Interoperability}

Graffiti applications are \emph{adversarially interoperable} with other Graffiti
applications, as per Requirement~\ref{requirements:adversarial-interop}.
However, Graffiti applications do not interoperate with \emph{external}
applications or protocols.
One tool that does is Gobo, which unifies feeds from Reddit, Bluesky,
Mastodon, and, previously, Twitter,
and allows users to customize the algorithms that sort their unified feed~\cite{gobo}.
Matrix also provides ``bridges'' to itself from messaging platforms like
Discord, Slack, and Messenger
\footnote{
    \url{https://matrix.org/ecosystem/bridges/}
}.
It may be possible to build bridges with other services either into or out of
Graffiti as a way to encourage adoption.

\subsection{Channel-Like Models}

Channels, as discussed in Section~\ref{concepts:channels}
are one of the novel aspects about the Graffiti API.
So far, we have discussed channels in relation to the publish-subscribe
pattern~\cite{pubsub}.
Here we mention two other concepts that relate to and shaped our
design of Graffiti's channels.

\subsubsection{Bidirectional Links}

On the World Wide Web, hyperlinks point in one direction, from one website to another.
However, hypertext systems like Xanadu~\cite{xanadu}
and more recently Roam\footnote{\url{https://roamresearch.com}}
and Notion\footnote{\url{https://www.notion.so}}
include \emph{bidirectional} links:
from one document you can see all the other documents that link to it.

Channels can be seen as a way to \emph{selectively} create bidirectional links.
A reply object points \emph{to} the object it is replying to.
Placing the reply in the channel represented by the original post's URL
creates a link in the opposite direction,
pointing \emph{from} the original post object \emph{to} the reply.
The ability for an object's creator to select which of its links should be bidirectional
is critical for managing context collapse.
The general Dexter hypertext model allows for mixed link types, but
as far as we can tell, it was never implemented~\cite{dexter}.

\subsubsection{Object Capabilities}
\label{related-work:ocaps}

%DK "Object capabilities" sounds so much better than "security by obscurity".   Can weintroduce object capabilities  early to describe graffiti's soft access control?
Object capability security is a security model where rights to perform an action
can be transferred to another user by giving the user a reference to that action~\cite{capabilitymyths}.
A channel name can be thought of as a ``read'' capability because knowledge
of a channel name allows an actor to read all public objects in that channel.

The Spritely Institute aims to encapsulate \emph{all}
security aspects of a social application with object capabilities~\cite{spritely}.
Capabilities can indeed be layered to create complex interactions,
including revocation of access.
However, many of these interactions require an agent acting on the user's behalf
that is always \emph{online}, similar to ActivityPub's server side effects.
Since we do not want developers to have to write server code,
we chose to give objects more familiar \texttt{allowed} lists
in addition to channels.

``Write'' capabilities are irrelevant in Graffiti given that total reification enables many coexisting permissions structures
to be built out of only self-writes.

\section{Conclusion and Future Work}

Graffiti seeks to enable a ``world where many worlds fit''~{\cite{escobarpluriverse}},
comprised of diverse social spaces that are each simultaneously \emph{personal}
and \emph{interoperable} with each other.
We introduce the concepts \emph{total reification} and \emph{channels}
to resolve the tension between these competing goals
of minimal constraint and interconnectedness.

In such an unconstrained environment,
our running hypothesis is
that the social pressures that currently work
to consolidate people on a handful of massive platforms
will similarly nudge users of Graffiti toward interoperability.
Specifically, we predict that natural \emph{folksonomies}~{\cite{folksonomy}} will
develop around the usage of channels and object schemas, much as they have with hashtags.
It remains to be seen whether this hypothesis will hold in wide deployment,
or whether there will be instances where too much freedom leads to
``expression breakdowns''~{\cite{expressionbreakdowns}}
that require additional layers of standardization.

Our work illustrates that by placing an \emph{API}
at the right level of the stack, we can separate out the distinct concerns
of building a usable application development environment
from concerns of building a secure, scalable, and decentralized infrastructure.
We demonstrate this by developing a Vue plugin and a series
of applications \emph{above} the API, independently from the
two decentralized implementations we present \emph{below}.
However, this hourglass design assumes the API in the middle is static,
which begs the question: is our API sufficient?

Our case studies span the Form-From taxonomy of social media~{\cite{formfrom}},
by demonstrating commons (Wikiffiti), spaces (Parallax, The Glue Factory), and networks (Glitter),
as well as both threaded (Glitter) and unthreaded (Parallax) interactions.
Channels and allowed lists respectively capture the two major
forms of context management available in social applications:
communication toward both abstract and targeted imagined audiences~{\cite{imaginedaudience}},
and communication toward a specific audience through ``visibility control''~{\cite{visibilitycontrol}} affordances.
Graffiti interactions can be ephemeral or persistent, real-time or asynchronous,
and moderated with practically any rule set.

However, while the Graffiti API theoretically provides the necessary
affordances for retrieving social data of interest,
there are cases where our client-side approach inevitably breaks down.
For example, it is unnecessarily difficult for a consumer device to
aggregate millions of reified likes on a celebrity's post into a single count, and
downright impossible for such a device to
process millions of videos to construct an analog of TikTok's For You feed.
For these cases, future work should develop
a parallel API for offloading such
computation to third-party services.
Similar care should be taken
in the design of that API to prevent lock-in.

Lastly, we framed the motivation behind Graffiti in terms of
\emph{individual} and \emph{community}-centric concerns,
specifically, the desire to personalize social spaces while remaining connected to others.
Our work does not directly address \emph{society}-level concerns,
such as disinformation and polarization.
Our hope, in line with predictions made by~{\cite{threeleggedstool}}, is that by democratizing
the design of communication environments---as Graffiti works to do---we
will see the bottom-up emergence of healthier social interactions within those environments,
at both community and society scales.
We look forward to continuing to work toward that vision.

\begin{acks}
We thank Caleb Malchik, Spencer Lane, Adam Kohan,
Joan Feigenbaum, and Ethan Zuckerman of the ``loyal clients'' group
for their regular feedback and thoughtful discussions.
We are also grateful to Josh Pollock and Leah Namissa Rosenbloom for their
comments on earlier drafts, to the members of Horse House
for being early adopters of Graffiti,
and to Susan Henderson for a final read-through.
Finally, we thank the anonymous reviewers for their constructive input.

\end{acks}

\bibliography{graffiti}

\appendix
\section{More Channel Name Constructions}
\label{appendix:more-channel-names}

Graffiti channels, as described in
Section~\ref{concepts:channels},
are designed to allow for new contextual meanings to evolve according to a folksonomy~\cite{folksonomy}.
In Table~\ref{concepts:channels-and-applications},
we listed some of those possible meanings.
Here, we expand on how continuous
contexts, like locations, can be represented with
discrete channel names and how channel names
can be combined to create new
intersectional meanings.

There are many standard models for representing
discrete physical \emph{areas} that could be co-opted to construct channel
names for location-centric contexts. These include
human-generated divisions, like zip codes,
and automatic divisions, like what3words\footnote{
    \url{https://what3words.com}
}, a tiling of the earth into three-meter squares.
The construction chosen for a particular application will
depend on that application's purpose,
its ``reach'' needs, and its privacy needs.
For example, a civic discussion application
might benefit from using voting precincts and wards
to construct channel names.
A classifieds application, like Craigslist, may benefit from posting
advertisement objects to channels representing multiple \emph{scales}
in a ``resolution pyramid,'' to encourage discoverability,
using a hierarchical area standard such as H3\footnote{
    \url{https://h3geo.org/}
}.
A location-based dating application, like Tinder, may intentionally
post profile objects only to coarse area channels to prevent stalking.
Conversely, when exact location is important,
as is the case for geocaching, an application could post the geocache
object to several coarse area-based channels to allow users
to search for geocaches nearby, but include
the exact cache location as an object \emph{property}
so that receiving applications can mark the cache on a map.

Channels can also be concatenated to produce new, more specific channels,
which may help to prevent context collapse in overloaded domains.
For example, a Tinder-like application may post profile objects
to channels with names of the form
\texttt{dating+\allowbreak{}zip:\allowbreak{}12345} rather than
\texttt{zip:\allowbreak{}12345} so that dating profiles
do not accidentally show up to neighbors browsing a local
discussion platform, like Nextdoor.
Concatenation may also happen in natural language.
For example, on Reddit,
programming-related memes are posted to \texttt{r/\allowbreak{}Programmer\allowbreak{}Humor}
rather than \texttt{r/\allowbreak{}programming+\allowbreak{}r/funny}.

\section{Announce Protocol}
\label{appendix:announce}

The remote implementation, described in Section~\ref{above-and-below:remote-protocol},
is currently built with a \emph{registry} and applications must
call \texttt{discover} on all the servers in the registry for
each high-level \texttt{discover} call.
This has the potential to be inefficient if the application queries
servers that have no objects, and also adds friction to participation,
as each new server needs to be ``registered'' to become discoverable
to the rest of the network.

Alternatively, it is possible to use an ``announce'' protocol,
which performs a similar service to that of the ``tracker'' used in
the commodity storage implementation (see Section~\ref{above-and-below:commodity-storage-protocol}). However, the announce protocol is built on top of the isolated Graffiti
API that is exposed by each remote implementation server, without the need for a separate
tracker service. Building on top of the Graffiti API also
allows the announce protocol to flexibly evolve as
new object properties are introduced.

The announce protocol assumes that the network is distributed like other
federated services, like Email or Mastodon, with a small handful of
massive ``well-known'' servers, like GMail or mastodon.social, and a long
tail of smaller, possibly self-hosted servers~\cite{mastodonchallenges}.
When an actor publishes an object to one such small server,
their client also publishes an \texttt{"AnnounceServer"}
activity, pointing to their small server, on several of the well-known
servers:
\begin{minted}{javascript}
{
  activity: "AnnounceServer",
  target: "https://my-small-server.com"
}
\end{minted}
This announcement must be crossposted to all the channels
that the actor posts to on the small server.
If there are channels where the actor has only posted
access-controlled objects, they must post additional announcement
objects with appropriate \texttt{allowed} lists.

To fulfill a high-level \texttt{discover} request,
the client must first \texttt{discover} for
\texttt{"AnnounceServer"} activities in the channels of
the high-level request
on each of the well-known servers.
Then, the client can \texttt{discover} for objects
matching the high-level channels \emph{and} schema on each of the targeted servers.
The protocol will not miss objects, so long as there
is an overlap between the retrieving and posting actors' well-known
servers lists.
The protocol will also not make requests to any server
that does not have objects in the channels of interest.
The protocol could be updated to also include object
\emph{schemas} in the announcements, as an additional
filter to prevent querying unnecessary servers.
This \texttt{"schemas"} property and other properties
can be flexibly added to \texttt{"AnnounceServer"} activities
in a folksonomic way, like any other Graffiti objects.

The announce protocol is not applicable to the commodity storage implementation
because commodity storage servers do not allow clients to
query multiple actors' data at once. A tracker is necessary
for the commodity storage implementation.

One downside of the announce protocol is that it leaks all
channel names to the well-known servers.
This may not be desirable to an individual or community that
decides to self-host specifically for privacy purposes.
This can be resolved by posting announcements to the channels
named by the \emph{hashes} of the actual channel names.
Alternatively, if the remote servers become end-to-end-encrypted,
as mentioned in Section~\ref{above-and-below:below:future-work},
this would no longer be an issue.

\end{document}